# Title: Large increases in public R&D investment are needed to avoid declines of US agricultural productivity

**Authors:** Ariel Ortiz-Bobea[1,2]*, Robert G. Chambers[3], Yurou He[1], David B. Lobell[4]

**Affiliations:**

[1]Charles H. Dyson School of Applied Economics and Management, Cornell University; Ithaca, NY, USA.

[2]Jeb E. Brooks School of Public Policy, Cornell University; Ithaca, NY, USA.

[3]Department of Agricultural and Resource Economics, University of Maryland, College Park; College Park, MD, USA.

[4]Department of Earth System Science and Center on Food Security and the Environment, Stanford University; Stanford, CA, USA.

*Corresponding author. Email: ao332@cornell.edu

**Abstract:**

Increasing agricultural productivity is a gradual process with significant time lags between research and development (R&D) investment and the resulting gains. We estimate the response of US agricultural Total Factor Productivity (TFP) to both R&D investment and weather, and quantify the public R&D spending required to offset the emerging impacts of climate change. We find that offsetting the climate-induced productivity slowdown by 2050 alone requires a sustained public R&D spending growth of 5.2–7.8% per year over 2021–2050. This amounts to an additional $208–$434B investment over this period. These are substantial requirements comparable to the public R&D spending growth that followed the two World Wars.

**Main Text:**

US agriculture is among the world's most productive due in large part to past public research and development (R&D) (*1,2*). Higher productivity increases farm profitability while reducing the sector's environmental impact and food prices (*3–5*). Despite substantial productivity gains, US agriculture remains acutely vulnerable to extreme weather (*6,7*). Recent climate trends have been detrimental to global agriculture (*8,9*) and there are signs both US and global agricultural productivity growth are slowing (*10,11*). Given the delayed benefits of R&D investments, it is pressing to assess US agriculture's R&D needs in a rapidly changing climate.

We quantify the public R&D spending growth needed to offset the emerging impacts of climate change on US agricultural productivity. We first estimate econometric models quantifying the historical effect of R&D spending and weather fluctuations on national agricultural Total Factor Productivity (TFP). TFP measures productivity and reflects the ratio of all measured agricultural outputs to inputs (*12*). We use these estimates to simulate how climate change would affect agricultural TFP via changes in counterfactual weather trajectories since 1950 based on climate model output from the 6$^{th}$ Coupled Intercomparison Project Model (CMIP6) (*13*). We then solve for the R&D spending needed to offset these emerging climate change impacts on TFP by certain target dates. Our estimates quantify the additional R&D spending required to avoid declines in US agricultural productivity under climate change with stagnant public R&D spending.

Our primary measures of national agricultural TFP and R&D expenditures come from official USDA-ERS statistics which we complement with other sources (*14*) (Fig. S1). We use annual public agricultural R&D expenditures to construct a stock of knowledge that explains the observed growth in agricultural TFP (Fig. 1A, 1F). We aggregate daily gridded weather datasets (*15,16*) to the national level based on various land cover weights (*17*) (Fig. S2) to explain year-to-year fluctuations in agricultural TFP. We then conduct counterfactual simulations based on downscaled bias-corrected data from CMIP6 over 1950–2100 (*18*) (Figs. S3–S5). We describe our data and methods in greater detail in the SM.

**Historical contribution of R&D on agricultural TFP**

Many studies analyze the link between R&D spending and agricultural productivity (*19–21*). Universities, government agencies, and private agricultural firms conduct agricultural R&D and translate the resulting knowledge into new technologies to be adopted by farmers with the support of extension services. This process is notoriously slow (*22*). Over time, we observe farmers producing more relative to the inputs they use, which manifests as TFP growth.

We propose a simple empirical model of how public agricultural R&D translates into TFP. Our baseline model takes the form:

$$\log A_t = \beta_0 + \beta_1 \log S_t + f(Z_t) + \epsilon_t \quad (1)$$

where $A_t$ is national agricultural TFP in year $t$, $S_t$ is an R&D or knowledge stock resulting from past public R&D activities, $Z_t$ is a vector of weather variables and $\epsilon_t$ is an error term. Following conventional practice, we model $S_t$ as a finite accumulation of knowledge flows proxied by public R&D expenditures $RD_t$ (*20*). Past R&D spending contributes for up to 50 years before reaching obsolescence, so that $S_t = \sum_{l=0}^{49} \gamma_l RD_{t-l}$ with research lags $\sum_{l=0}^{49} \gamma_l = 1$.

This framework has several key implications. First, it suggests that TFP is primarily influenced by *public* agricultural R&D. Private R&D is primarily aimed at technologies that are embodied in the inputs so they can be sold to farmers. The official TFP estimates we use account for inputs that reflect such "quality" changes. As a result, TFP reflects disembodied technical change that arguably should stem primarily from public R&D. We nonetheless revisit this assumption empirically. Second, the stock of knowledge $S_t$ affects the level of TFP directly, not its growth (*22*). This means that constant R&D spending eventually leads to a stabilization of $S_t$ and thus to TFP stagnation. This reflects the need for "maintenance" research to keep agricultural productivity from declining due to everchanging pest and weed pressures (*23*).

Previous work finds the benefits of agricultural R&D rise and then fall over time (*24*). We follow (*1,25*) and represent the research lags $\gamma_l$ with a gamma distribution. This allows us to consider a wide range of lag weight structures with only 2 parameters (Fig. S6). We select a baseline model based on the research lags that best predict the historical levels of TFP.

Public agricultural R&D expenditures underwent a steady rise throughout most of the 20[th] century, but slowed starting in 1970, stagnated around 2000, and then declined in the last decade (Fig. 1A) (*26*). Our model of the TFP response to R&D spending suggest the research lags peak around the 30[th] year (Fig. 1B–F). Notably, our model reproduces the recent slowdown in TFP growth (*27,28*), which results from the slowdown of the knowledge stock $S_t$ as seen in Fig. 1A. This slowdown in $S_t$, in turn, follows from the slowdown of R&D spending that began in the 1970s. The 30-year lag between a slowdown in R&D spending and a subsequent slowdown of the growth in the knowledge stock $S_t$ (and thus of TFP growth) has critical implications for future TFP growth. In particular, the ramifications of the stagnation and decline in investment over the past few decades have, for the most part, not yet materialized.

A common concern when estimating time-series models like equation (1) is the presence of non-stationary variables, in this case TFP and $S_t$. Two non-stationary variables can easily appear to be statistically related when they are not. To avoid spurious inference, it is common to rely on higher-frequency or de-trended variation of the data for identification. This can take the form of a first-difference model or the use of a time-series filter to remove the trends that underpin the spurious relationship (e.g. *29*). However, we know based on theory that $S_t$ is the primary driver of TFP. This conceptual clarity alleviates concerns regarding the nature of the relationship. However, it does not resolve issues related to the omission of other R&D activities that could bias the magnitude of the relationship.

A key challenge in analyzing the return of R&D on TFP is the presence of spillovers (*30–32*). For instance, agricultural TFP can grow due to spill-ins of agricultural R&D from other regions or from non-agricultural R&D. If these other sources of R&D are correlated with public agricultural R&D, then not controlling for them in equation (1) could overstate the returns of public agricultural R&D. Our baseline model yields elasticities of the knowledge stock $\beta_1$ around 0.5 (Fig. 1D, Table S1). This means that increasing TFP by 1% requires about a 2% increase in the knowledge stock. This is slightly higher than other studies seeking to isolate the effect of public R&D specifically (with $\beta_1 \sim 0.3$, e.g. *33*). Table S1 shows that controlling for a linear time trend leads to smaller estimates aligned with previous work. However, applying a more sophisticated time-series filter proposed in (*34*) yields similar (but noisier) estimates to our baseline model (Table S1, columns 1 vs 4). This suggests that spill-ins from other countries or sectors might be limited. We thus favor a baseline model estimated in levels to avoid the additional noise introduced when relying on the high-frequency variation of slow-moving variables.

We consider several variations of this model. First, we find that computing a knowledge stock from both public and private agricultural R&D points to qualitatively similar elasticities (Table S2). This is consistent with (*35*) who find an ambiguous and noisy effect of private R&D on TFP, and the idea that technological change from private R&D is embodied in inputs and thus not reflected in TFP. We also find slightly larger estimates of the elasticity of the research stock when we consider an alternative agricultural data source (Table S3). Finally, we find that excluding weather in all these models has little bearing on the elasticity (column 2 in Tables S1–S3).

A potential concern is that our estimate of the returns of R&D may be biased upward due to the omission of rising $CO_2$ fertilization which boosts photosynthesis, especially of C3 crops. We explore this possibility by including summer $CO_2$ concentration as an additional control (Table S4). We find that our baseline estimate remains largely unchanged (columns 1–3). We also explore a specification with a Hamilton trend and obtain a large but imprecise estimate for the elasticity of the knowledge stock when we control for $CO_2$ concentration (columns 4–6). We find no evidence that $CO_2$ fertilization is affecting our estimates of R&D returns. We also find that estimates of the contribution of $CO_2$ to TFP is imprecise, in line with previous work on crop yields (*36*).

**Historical effect of weather on agricultural TFP**

Estimating the effect of weather on TFP from equation (1) harnesses variation around the predicted contribution of the knowledge stock (Fig. 1F). This approach may not control fully for data trends that could inadvertently make our estimates less precise. We therefore estimate weather effects separately from R&D effects in an auxiliary regression that controls for trends using the Hamilton time-series filter. Thus, weather effects are estimated around a flexible trend. To ensure that we capture the covariance between R&D and weather effects, we sample observations in the same way in each model when performing the block bootstrap procedure described in the SM.

We find that exposure to temperatures exceeding 25°C reduce TFP (Fig. 2A), with a strong nonlinear effect of temperature similar to previous work on crop yields (*37,38*). Our model explains a significant fraction of TFP variations (Fig. 2B), with warmer conditions associated with lower TFP. We also consider how uniform changes in annual temperature and precipitation would reduce TFP (Fig. 2C–D), finding that total effects are dominated by temperature, again consistent with prior work (*38–40*) (Tables S5 and S6). A 3°C warming reduces TFP by more than 10%. We also confirm weather effects are noisier when estimated around the predicted value of the knowledge stock (Fig. S7).

Our baseline model is based on weather variables aggregated over the entire calendar year. We implement an approach proposed in (*41*) to recover the underlying seasonality in the data. In general, we find that aggregating weather variables over the summer months improves TFP predictions out of sample (Fig. S8). But we also find that models based on temperature exposure bins still perform well when variables are aggregated over the entire year. And tailoring weather variables to the best-fitting season does not substantially change the weather sensitivities of TFP relative to our baseline model (Fig. S9). We thus favor a baseline model based on annual temperature bins and precipitation.

We consider multiple variations of the model. First, we consider alternative specifications that rely on linear and quadratic time trends rather than on a the time-series filter (Fig. S10–S11). We find those estimates to be less precise than our baseline model. Second, we consider models based on weather data aggregated using alternative spatial weights (Fig. S2). We find that estimates based on equal spatial weights are smaller and less precise (Fig. S12) but that those based on cropland and pasture weights are comparable to our baseline model that accounts for the value of production (Fig. S13–S14). We also find comparable estimates when we consider an alternative weather dataset (Fig. S15).

**Projected impact of climate change on agricultural TFP**

We compute the impact of climate change on TFP by simulating the effect of weather trajectories from counterfactual climate simulations (Fig. 3). This approach combines the statistical uncertainty from the econometric model, reflected in the estimated TFP sensitivities to weather, with the climate uncertainty stemming from alternative General Circulation Models (GCMs) along the 4 main Shared Socioeconomic Pathways (SSPs) in CMIP6. The projected TFP impact by 2050 ranges from –7.8% to –13.1% depending on the SSP (Table S8). As expected, the range of impacts expands from –8.4% to –35.2% across SSPs by 2100 (Table S10). Note that projected impacts by 2025 already appear somewhat negative at around –5.6% for SSP2-4.5 (with a 95% C.I. of –11.2% and +0.2%) (Table S7). This indicates that climate change is projected to be having a detrimental effect on US agricultural productivity, although the impact of recent climate trends is not yet clear (Fig. S16). The bottom panels of Fig. 3 show impacts of climate change on TFP separately by GCM, thus separating the econometric and climate uncertainties. These estimates are also presented separately by GCM and for several target years in tables S7–S10.

A potential concern is that historical TFP sensitivities to weather could overstate the future sensitivities. Indeed, climate-smart agricultural practices and technologies could improve climate resilience over time (embodied technical change). We do not consider this a major caveat for several reasons. First, there is no clear evidence that US agriculture has been adapting to recent climate trends (*42*). Second, various studies document that US agriculture has become more sensitive to high temperatures over time (*6,7,43*). Third, while improving climate resilience is plausible, technological change takes time due to long research lags, as highlighted by our study.

Our projected climate change damages on US agricultural TFP are smaller than in (*7*). We attribute this difference to the way we model the effect of weather on TFP. In our model, weather affects the level of TFP, not TFP growth. That is, a weather shock only has a transitory effect on TFP and does not affect its future trajectory. This assumption could understate the damage of climate change even if small growth effects are present but are ignored.

**Projected R&D spending needs to offset climate change damages**

To estimate the R&D spending needed to offset climate change damages, we first solve for the change in the knowledge stock $S_t$ that fully offsets climate change impacts on TFP over time (Fig. S17A–D and Table S11). Estimates indicate that the knowledge stock needs to be between 16.7% to 28.2% larger by 2050 and between 16.0% and 151.0% larger by 2100 depending on the SSP scenario. These estimates reflect both the statistical uncertainty from the econometric model as well as the climate uncertainty reflected in CMIP6.

To translate these relative changes in $S_t$ to monetary values we need to assume a counterfactual level of $S_t$ without climate change. We assume that the counterfactual $S_t$ without climate change would remain constant at its 2020 level until the end of the century. Figs. 4A and S17 (panels E–H) and the bottom of Table S11 thus show the monetary changes in $S_t$ needed to keep TFP from declining over time in the presence of a climate-induced slowdown in TFP. The additional size of $S_t$ ranges from about $0.9B to $1.6B by 2050 and $0.8B to $9.4B by 2100 depending on the SSP scenario. Note that permanently increasing $S_t$ by $1B, requires a steady-state increase of $1B per year in R&D spending.

There is an infinite number of R&D spending pathways to compensate the effect of climate change on TFP by a certain date. This relates to how R&D spending contributes sequentially to the knowledge stock $S_t$ through long research lags. For simplicity, we search for R&D spending pathways with fixed annual growth rates, starting in 2021. Intuitively, reaching a certain change in $S_t$ requires faster annual spending growth if the target date is closer in time. We solve for these constant-growth spending trajectories for various target years, and we illustrate how we conduct the stock-to-spending mapping in Fig. S18.

To offset climate change damages on TFP by 2050, we estimate that annual R&D spending would need to grow by 5.2% to 7.8% per year starting in 2021 depending on the SSP scenario (Figure 4 and Table S12). For context, Fig. 4B also showcases historical average annual R&D spending growth rates by decade. While our estimates remain imprecise, our central estimates point to R&D spending growth levels that rival those seen following the First and Second World Wars. Monetarily, these needs amount to an *additional* $207.6 to $433.7B in public R&D spending between 2021 and 2050 above a baseline investment of ~$5B per year observed in 2020 (Table S13). For comparison, the recent Inflation Reduction Act provided an additional $19.5B over a 5-year period to the USDA to support farm conservations practices to mitigate climate change.

Public R&D spending needs by 2050 are large for 2 reasons. First, the climate-induced slowdown of US agricultural TFP is imminent if not ongoing. Fig. 3 shows that projected climate change damages by 2025 are close to being negative. Second, we find that research lags are long. Specifically, Fig. 1 shows that the impact of R&D spending peaks around the 30[th] year. In summary, imminent climate impacts along slow R&D calls for large offsetting investments in the next 30 years.

Naturally, the growth in public R&D spending to offset climate change impacts on TFP is smaller when considering target dates beyond 2050. Fig. S17 (panels I–L) and Table S12 show the annual R&D spending growth needed under alternative SSPs and target dates. As expected, spending requirements are closer across SSPs for target dates that are near because TFP impacts are relatively similar across SSPs in the next few decades (Fig. 3). Because spending growth compounds over time, the requirements to offset damages by 2100 are smaller on a yearly basis than those needed for 2050. However, the needs remain substantial. For instance, offsetting climate change damages on TFP by 2100 requires an annual R&D spending growth of 0.3% to 2.2% per year starting in 2021 depending on the SSP scenario (Table S12). Note that public R&D spending has grown by just 0.5% per year over 1970–2020. These needs amount to an *additional* $54.6 to $679.9B in public R&D spending between 2021 and 2100 above a baseline investment of ~$5B per year (Table S13). Fig. S19 illustrates how future TFP levels will fundamentally depend on R&D spending and the severity of climate change.

We should highlight that the public R&D spending needs we outline here simply offset the slowing effect of climate change on TFP. This can be interpreted as the R&D needed to avoid TFP from declining given a fixed baseline R&D investment. Our framework suggests that simply meeting these targets would result in a stagnation of TFP. Thus, maintaining historical level of TFP growth will require even higher increases in public R&D spending.

**Acknowledgments:**

**Funding:**

US Department of Agriculture NIFA grants 2021-67023-34492 (AOB, RGC)

US Department of Agriculture NIFA Hatch/Multi State project 7003346 (AOB)

**Author contributions:**

Conceptualization: AOB, RGC, DBL

Methodology: AOB, DBL

Investigation: AOB, YH

Visualization: AOB

Funding acquisition: AOB, RGC, DBL

Project administration: AOB

Supervision: AOB

Writing – original draft: AOB

Writing – review & editing: AOB, RGC, YH, DBL

**Competing interests:** Authors declare that they have no competing interests.

**Data and materials availability:** Data and code necessary to reproduce the analysis are permanently available from the Cornell Center for Social Sciences (https://doi.org/10.6077/q59w-tj45). We do not share some of the source data due to size and licensing restrictions. But all necessary secondary data files and script files used to convert source data to secondary data are also provided.

**Supplementary Materials**

Data and Methods

Figs. S1 to S20

Tables S1 to S13

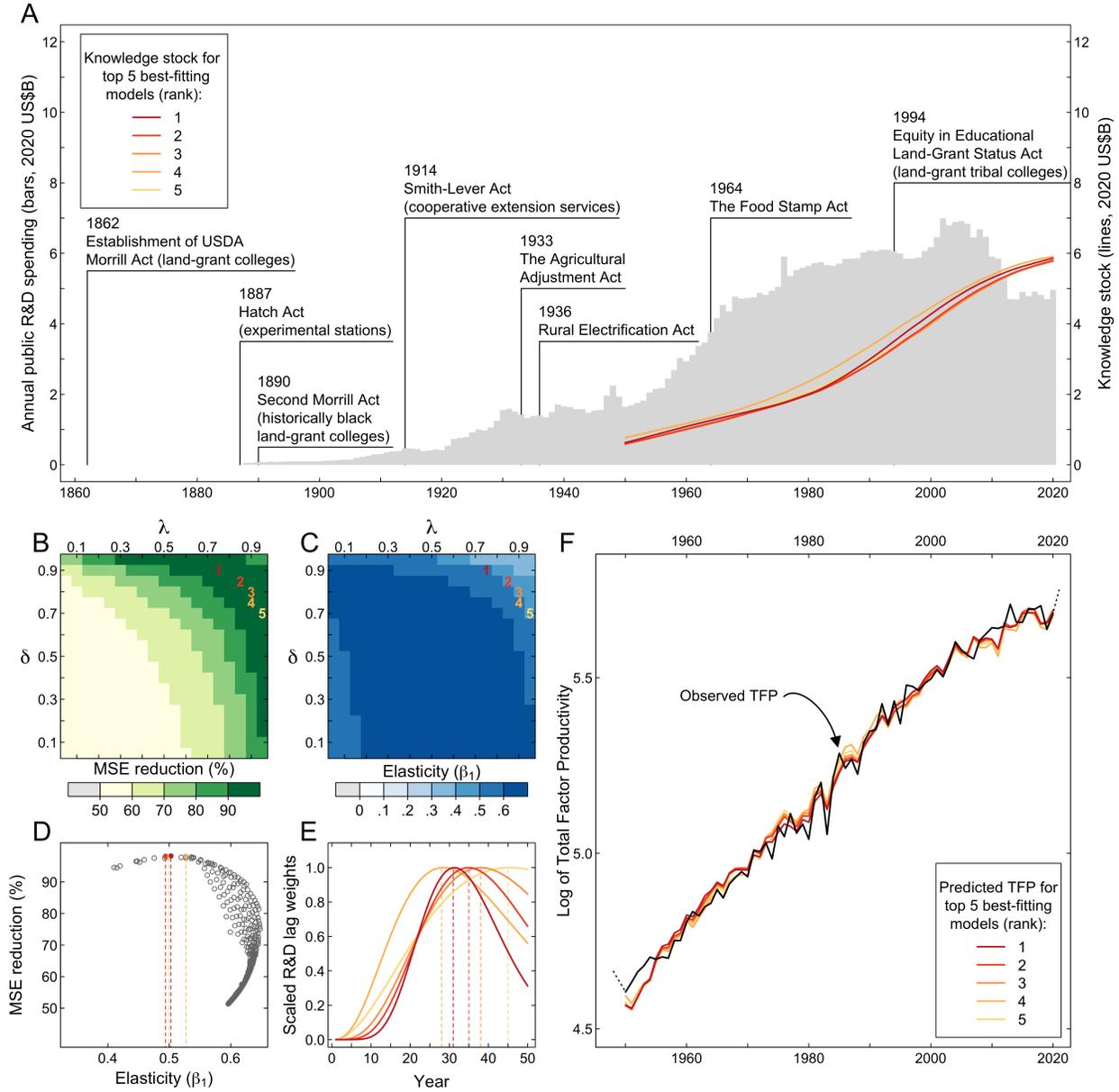

**Fig. 1. Historical contribution of US public agricultural R&D expenditures on national agricultural Total Factor Productivity (TFP).** (**A**) The evolution of agricultural public R&D spending during 1888–2020 along with several historical milestones of US agriculture. The colored lines correspond to stock of knowledge $S_t$ corresponding to the top 5 best-fitting model specifications identified in panels B-E. (**B**) Reductions in out-of-sample MSE of models with alternative specifications of the knowledge stock $S_t$ relative to a model without $S_t$. We conduct the grid search over the 2 parameters ($\delta$ and $\lambda$) of the gamma distribution. The top 5 best-fitting models are highlighted with their rank (1–5). (**C**) Value of the estimated elasticity of the knowledge stock $\beta_1$ over the gamma distribution parameter space. The top 5 best-fitting models are also highlighted. (**D**) Scatterplot of panels C and D highlighting the top 5 best-fitting. (**E**) Shape of the research lags corresponding to the top 5 best-fitting models. (**F**) The predicted value of log TFP according to the top 5 best-fitting models identified in previous panels. The black line represents the observed TFP trajectory. Dashed black lines correspond to TFP data excluded due to R&D or weather data limitations (see SM).

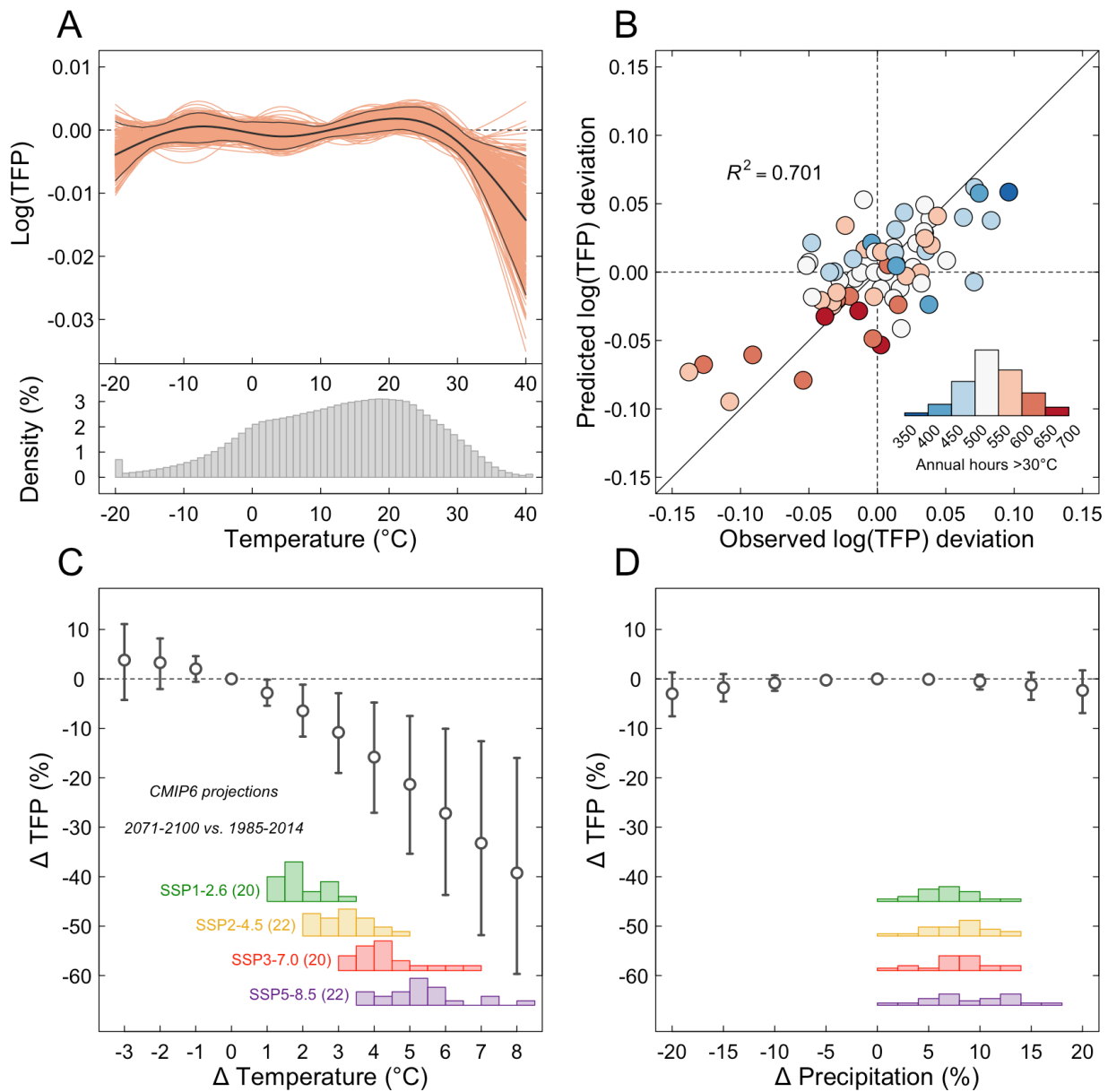

**Fig. 2. Historical impact of weather fluctuations on US agricultural Total Factor Productivity.** (**A**) Effect of additional annual exposure to various levels of temperature on TFP. The various red lines represent 500 bootstrapped estimates. Black lines represent 95% pointwise confidence intervals. (**B**) Scatterplot of observed and predicted TFP deviations from a linear trend. (**C**) Impact of uniform changes in annual temperature on TFP. (**D**) Impact of uniform changes in annual precipitation on TFP. For reference, panels C and D show the distribution of national changes in annual temperature and precipitation from GCMs in CMIP6 between the end of the century (2071–2100) and a historical reference period (1985–2014).

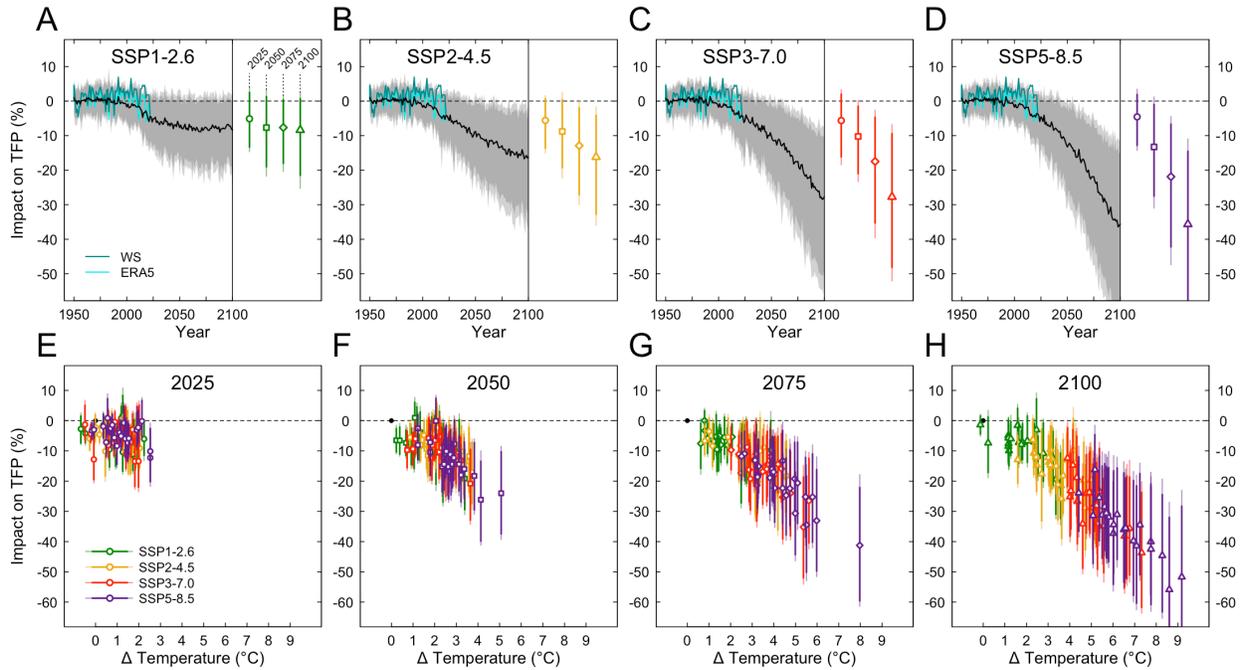

**Fig. 3. Impact of climate change on US agricultural TFP.** (**A**–**D**) The top row of panels shows the impact of climate change over time for the 4 main SSPs since 1950. As highlighted in panel A, we also show estimates for specific target years: 2025, 2050, 2075 and 2100. These estimates reflect both the econometric uncertainty reflects in the estimated parameters and the climate uncertainty stemming from different GCMs in CMIP6. (**E**–**H**) These panels show the impacts of climate change since 1950 for each GCM separately, as a function of annual temperature change and at different target years. The intervals around each estimate represent the econometric uncertainty.

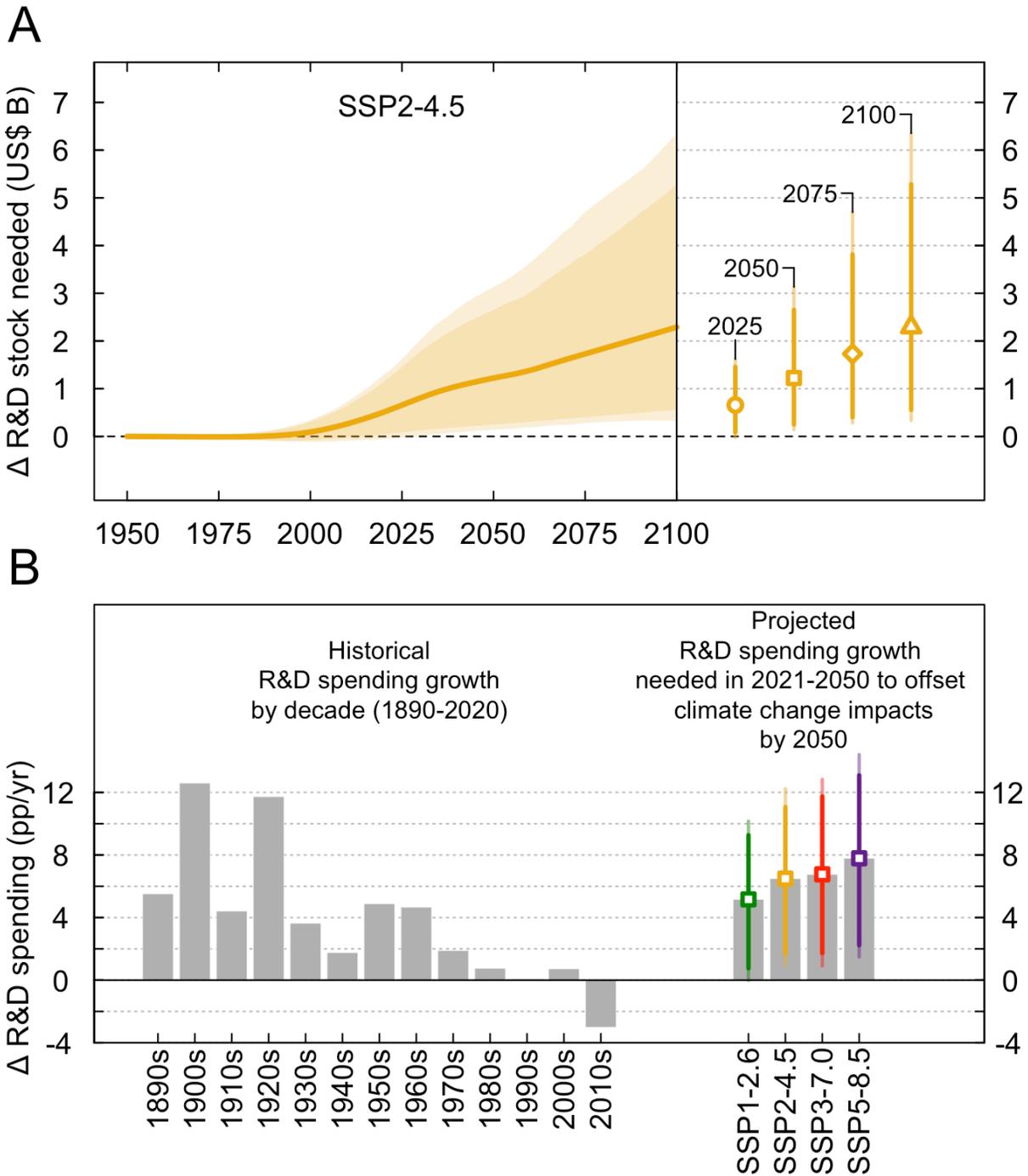

**Fig. 4. R&D stock and spending growth needed to offset climate change impacts on US agricultural TFP.** (**A**) Change in the value of the knowledge stock $\Delta S_t$ that would offset the impact of climate change on US agricultural TFP under SSP2-4.5. The panel also shows estimates for several reference years: 20025, 2050, 2075 and 2100. (**B**) Shows the observed growth in public agricultural R&D spending by decade (left) along with the R&D spending growth needed to offset climate change impacts by 2050. The error bands reflect both econometric uncertainty and climate uncertainty from the various GCMs in CMIP6.

# Supplementary Materials for

# Large increases in public R&D investment are needed to avoid declines of US agricultural productivity

Ariel Ortiz-Bobea, Robert G. Chambers, Yurou He, David B. Lobell

Corresponding author: ao332@cornell.edu

**The PDF file includes:**

Data and Methods

Figs. S1 to S20

Tables S1 to S13

**Data and Methods**

Data

*Agricultural data*

Our study primarily relies on official estimates of Total Factor Productivity (TFP) for the US agricultural sector from USDA-ERS. The 2024 version of the dataset provides annual estimates of US agricultural TFP for 1948–2021. Details about the construction of these estimates are provided on the USDA-ERS website.[1] Some of our analysis also relies on an alternative dataset (1949–2007) from the International Science & Technology Practice & Policy (InSTePP) group at the University of Minnesota. More details about this dataset can be found on the InSTePP website.[2] Both data sources account for quality adjustments of inputs over time. This implies these TFP estimates aim to reflect disembodied technical change. We expand on this later.

We also rely on a long time series (1888–2020) of US public agricultural expenditures in research and development (R&D). We combine 2 separate data sources. We obtain data for 1888-1969 from (*44*) and data for 1970–2020 from USDA-ERS[3]. Expenditures are deflated to constant 2020 US$ using the NIH Biomedical R&D Price Index. Public agricultural R&D expenditures includes spending from 1). all USDA agencies (except the US Forest Service), 2). State Agricultural Experimental Stations (SAES) and 3). State Schools of Veterinary Medicine.

We also rely on an alternative public agricultural R&D dataset from InSTePP (1890–2007) (*45*)[2]. In addition to the spending categories listed above, this dataset also incorporates expenditures related to extension. As a result, the total public expenditures in this dataset slightly exceed those from the public agricultural R&D expenditures in the dataset described above (see Fig. S1B).

One of our analyses also relies on "total" agricultural R&D expenditures that include both public and private sources (Fig. S1B and Table S2). We obtain private agricultural R&D expenditures from 2 sources. We obtain data for 1920–1959 from (*44*) and data for 1960–2014 from USDA-ERS[3]. These private R&D expenditures relate to various types of agricultural inputs sold to farmers, and include: 1). crop breeding, 2). chemicals for crop protection, 3). synthetic fertilizers, 4). machinery, 5). animal health, 6). animal breeding and 7). animal nutrition.

---

[1] https://www.ers.usda.gov/data-products/agricultural-productivity-in-the-u-s/productivity-growth-in-u-s-agriculture/
[2] https://instepp.umn.edu/data/instepp-productivity-accounts
[3] https://www.ers.usda.gov/data-products/agricultural-and-food-research-and-development-expenditures-in-the-united-states/

Our analysis relies on detailed gridded climate data that we aggregate to the national level for the econometric analysis. We construct spatial aggregation weights based on the US Geological Survey (USGS)'s 2016 National Land Cover Database (NLCD) for the contiguous US (46)[4]. This is a 30-m resolution dataset with 16 land cover classifications. We capture cropland and pastures based on the "Cultivated Crops" and "Pasture/Hay" land cover classes, respectively. We compute the number of 30-m pixels in these land use categories within each grid cell of the gridded climate datasets we use. We illustrate the various types of aggregation weights we use in Fig. S2. Following (47), we aggregate the climate raster data using sparse matrices based on these aggregation weights. Finally, we also rely on a state-level agricultural output, input and TFP dataset from USDA-ERS (1960–2004). We rely on this dataset to scale agricultural land cover aggregations weights for weather data by production value (Fig. S2D). We do not rely on these state-level TFP estimates because we do not have access to the corresponding state-level R&D data.

*Historical climate data*

We primarily rely on a daily gridded weather maintained by Prof. Wolfram Schlenker (15). We refer to this as the "WS" dataset throughout the manuscript. This dataset provides daily minimum and maximum temperature and total precipitation over 1950–2022. The dataset is largely based on the PRISM dataset (48)[5]. Advantages of WS relative to the daily PRISM data include: 1- a longer temporal coverage (starts in 1950 instead of 1981 for PRISM) and 2- maintains a more stable set of underlying weather stations used in the spatial interpolation procedure. More details are provided on the dataset webpage.

We also rely on weather data from the ERA5 dataset of the European Centre for Medium-Range Weather Forecasts (ECMWF) (49)[6]. The ERA5 is a global hourly reanalysis dataset available since 1940. We download the data directly from the ECMWF servers using Python code that we provide with our reproduction package. We crop the data to the lower 48 US states and compute daily minimum and maximum temperature and total precipitation.

---

[4] https://www.mrlc.gov/data/nlcd-2016-land-cover-conus
[5] https://prism.oregonstate.edu
[6] http://cds.climate.copernicus.eu

*Modeled climate data*

Our study relies on output from General Circulation Models (GCMs) to simulate counterfactual changes in agricultural TFP. We rely on output from GCMs represented in the 6$^{th}$ Coupled Intercomparison Project Model (CMIP6) (*13*). We rely specifically rely on a daily version of this ensemble downscaled and bias-corrected by the Climate Impacts Lab to the ERA5 data (*50,51*). We select daily minimum and maximum temperature and total precipitation from 1950 to 2014 (historical simulation) and 2015–2100 for the 4 main Shared Socio-economic Pathways (SSPs) in CMIP6. We download the data Microsoft's Planetary Computer using Python code that we provide with our reproduction package. We crop the data to the lower 48 US states similarly to the ERA5 dataset.

*Climate data processing and variable construction*

A challenge with conducting an econometric analysis with a national time series is that weather predictors need to be aggregated to the same national scale. Aggregation either in space or time often leads to loss of information. Our aggregation procedures and criteria seek to minimize this information loss in various ways.

The first consideration is spatial aggregation. We aggregate gridded weather and climate datasets (WS, ERA5, CMIP6) based on spatial weights that represent agricultural land cover and production value. Fig. S2 shows 4 sets of aggregation weights. Figs. S3–S5 show key variables for these climate datasets under these 4 aggregation weights. The main analysis in the paper is based on cropland and pastures weights that are scaled by the state's value of production. As a result, cropland and pastures in California, a state with high-value specialty crops, exhibits relatively higher spatial weights, than areas in the Midwest which are more specialized in relatively less valuable field crops. In the main text we discuss that only the equal spatial weights appear to deteriorate model fit. Other aggregation strategies appear to provide similar results.

The second consideration is temporal aggregation. Most commodity-specific studies assume a relevant season for weather aggregation that roughly corresponds to the growing season in the region of interest. However, our analysis encompasses the entire US agricultural sector and thus reflects the sensitivities to weather fluctuations of all crops and livestock production. A simple strategy may be to aggregate weather information to the calendar year. A potential disadvantage with this approach is that temporal aggregation might conceal relevant within-year variation for the prediction of agricultural outcomes. We rely on an approach introduced in (*41*) that seeks to recover the underlying seasonality of the data when that seasonality is unknown. The approach consists of a grid search of all possible calendar periods within the year to identify the best-fitting seasons for a particular outcome. We conduct this analysis for the US national TFP data and show the findings in Fig. S8. As mentioned in the main text, we ultimately rely on variables aggregated to the entire calendar year rather than tailored variables to the best-fitting season, which do not point to substantially different findings.

A key feature of our analysis is the measure of temperature. We follow the approach in (*37*) and described in greater detail in (*47*) to compute measure of monthly exposure to 1°C intervals between –30 and +50°C. Computing these exposures requires information about the intra-daily distribution of temperature. We derive this by fitting a double sine curve passing through the minimum and maximum temperature of consecutive days and tracking exposure at 15-minute intervals. We perform this calculation for both historical (WS and ERA5) and CMIP6 datasets. The main goal of this approach is to avoid losing information about the occurrence of extreme temperature events when performing spatio-temporal aggregation. Indeed, models based on temperature bins are more forgiving than models based on average temperatures (see Fig. S8). Specifically, out-of-sample model predictions do not deteriorate as much when long aggregation windows (e.g. January–December) are selected. We later describe how we estimate non-linear effects of temperature exposure using a natural cubic spline.

Methods

*Conceptual framework*

We adopt a standard framework based on a Cobb-Douglas production function with constant returns to scale:

$$Y_t = \underbrace{a_t e^{f(Z_t)}}_{A_t} \prod_j X_{jt}^{\alpha_j} \tag{1}$$

where $Y_t$ is an aggregate output in year $t$, $X_{jt}$ is the quantity of the $j^{\text{th}}$ measured input, $\sum_j \alpha_j = 1$, and $A_t$ is Total Factor Productivity (TFP). Weather $Z_t$ is an exogenous random input vector that is not conventionally accounted for in official statistics and is thus subsumed in $A_t$. The function $f(\cdot)$ is plausibly a nonlinear function of $Z$. Re-arranging this equation and taking logs yields the definition that TFP growth equals the growth of output that cannot be explained by the growth of measured inputs:

$$\log A_t = \log a_t + f(Z_t) \equiv \log Y_t - \sum_{j=1} \log X_{jt}^{\alpha_j} \tag{2}$$

Our study quantifies the changes in R&D needed to avoid future declines in TFP. To clarify the scope of our analysis we first need to clarify the different ways technological progress can operate in this framework.

Technical change is said to be "embodied" (in the inputs) if it manifested through changes in the quality of the $X_j$ themselves. These quality adjustments can take different forms but apply to all inputs. For instance, workers can become more educated; the formulation of fertilizers can improve over time; machinery can become more advanced; crops and animal can be improved genetically, etc. We can characterize such quality adjustments as $X_{jt} = b_{jt} X_{jt}^*$, where $b_{jt}$ is a quality adjustment factor for the "conventional" input $X_{jt}^*$. In general, changes in input quality are captured through prices. For instance, educated workers command a higher wage, or improved seeds are more expensive. In a competitive market, input prices reflect the value of their marginal product. In other cases, however, input prices may be missing or distorted, in which case one can recover the underlying quality changes through a hedonic analysis (*52,53*). Note that $b_{jt}$ can be thought of as an "augmenting" factor, so this type of technical change can be interpreted as a form of *factor-augmenting* technical change.

Our framework and the official TFP statistics we rely upon are based on quality-adjusted inputs. TFP growth in our framework is thus equal to the growth of output minus the growth of measured *quality-adjusted* inputs. As a result, technological progress that is embodied in the inputs is not captured by our TFP measure. TFP captures (through $a_t$) what is commonly referred to as Hicks-neutral or "disembodied" technical change.

Our study focuses on *public* agricultural R&D spending needed to offset climate change impacts on TFP. One reason is that private R&D is primarily oriented towards generating embodied technical change which is not reflected in TFP. Private R&D firms improve the "quality" of inputs that are sold to farmers in the form of enhanced chemicals, seeds and machinery. These higher-quality inputs are sold at a price that reflects the quality of input services delivered, which does not affect TFP (as long as the quality-adjustment technique is correct). Public R&D on the other hand could lead to both embodied and disembodied technical change. As a result, we posit that the rise in TFP over time stems primarily from public R&D.

Following conventional practice, we assume that the non-weather component of TFP, $a_t$, is a function of a knowledge stock $S_t$ reflecting a distributed lag of past investments (or flows) in public R&D, $RD_t$:

$$\log A_t = \beta_0 + \beta_1 \log S_t + f(Z_t) \tag{3a}$$

$$S_t = \sum_{l=0}^{L-1} \gamma_l RD_{t-l} \tag{3b}$$

where $\beta_1$ is the elasticity of TFP relative to the knowledge stock and $L$ is the length of the so-called research lag. We later parametrize these functions in the empirical model.

We conceptualize the effect of climate change on TFP over time, $D_t$, as the effect of changing weather $Z_t$ relative to the effect of a reference climatology $\bar{Z}_0$:

$$D_t = \log A_t - \log A_0 = f(Z_t) - f(\bar{Z}_0) \tag{4}$$

We can derive the change in the knowledge stock $\Delta \log S_t^* = \log S_t^* - \log S_0$ that fully offsets $D_t$ to ensure ensures that TFP does not change:

$$\Delta \log S_t^* = -\frac{D_t}{\beta_1} \tag{5}$$

The more productive R&D is (larger $\beta_1$), the smaller the change in the knowledge stock $\Delta S_t$ needs to be to offset emerging climate change impacts $D_t$. Once $\Delta S_T$ is established for a given target year T, we can solve for sequences of annual public R&D expenditures, $RD_t$. We consider R&D spending paths with fixed growth rates between the year 2021 and the target year.

Our framework relies on the historical information of how the knowledge stock and weather affect TFP to conduct counterfactual simulation in the future. We address this concern in the main text.

*Estimating R&D effects on TFP*

Our econometric model follows from the equation (3a) and takes the general form:

$$\log A_t = \beta_0 + \beta_1 \log S_t + f(Z_t) + \epsilon_t \tag{6}$$

where $\epsilon_t$ is an error term that is serially correlated. In all models with account for such serial correlation with an overlapping block bootstrap (500 bootstraps) with blocks of 5 observations. We discuss the empirical parametrization of $f(\cdot)$ later.

Our national TFP data has a limited number of observations so estimating all the parameters in the distributed lag in equation (3b) is infeasible. We follow (*1*,*25*) and generate lag weights $\gamma_l$ using a gamma distribution. This allows us to generate a wide range of lag weight structures with only 2 parameters of this distribution ($\lambda$ and $\delta$, see Fig. S6). We conduct a grid search to find the best-fitting combination of these two parameters (Fig. 1B). We find $\lambda = 0.75$ and $\delta = 0.90$ is the best-fitting combination.

Table S1, column 1, shows a baseline estimate of 0.503 for $\beta_1$. Excluding weather as a control has little bearing on this estimate (column 2). Our estimate is slightly larger than previous reported estimates (e.g. *33*) that control for a linear time trend. Indeed, we find a smaller but less precise estimate when we control for a time trend (column 3). That model harnesses variation in TFP and the knowledge stock $S_t$ around a linear trend for identification. We also consider a model that controls for a Hamilton trend (*34*). The Hamilton filter was recently proposed as a strategy for removing the trend component of a time series. In our application, we include the 4 recent TFP lags to control for a flexible TFP trend (Fig. S20).

Table S2 replicates the analysis on Table S1 but with an alternative measure of the knowledge stock that incorporates both public and private R&D. We find that the new estimates are qualitatively similar to those in Table S1, although estimates accounting for a linear time trend appear no longer significant at conventional levels (column 3). Table S3 also replicates the analysis in Table S1 but with an alternative (but shorter) time series for both TFP and public R&D expenditures. Estimates for $\beta_1$ appear slightly higher but are no longer significant at conventional levels when we apply a Hamilton filter to this shorter dataset (column 4).

The concentration of carbon dioxide during the summer months of the northern hemisphere (June–August) rose from around 316 ppm in 1958 to about 415 ppm in 2020. Carbon dioxide is beneficial for photosynthesis, especially for C3 plants, which could have boosted productivity during the sample period. In Table S4 we explore whether including carbon dioxide as a control alters our baseline estimates. We find estimates remain largely unchanged when including carbon dioxide as a control (columns 1 vs 3 and 4 vs 6). Furthermore, while carbon dioxide on its own is positively correlated with the level of TFP (column 2), that association is no longer present when we control for the knowledge stock (column 2 vs 3) or when we control for a Hamilton trend (columns 5 and 6). While there is extensive evidence that carbon dioxide may be beneficial to crop productivity in controlled environments, we find that its contribution in this aggregate setting is too imprecise to capture.

*Estimating weather effects on TFP*

Our study considers the effect of temperature and precipitation on TFP. We estimate these weather effects with a model of the form:

$$\log A_t = \beta_0 + g(t) + \sum_{j=1}^{J}\sum_{k=1}^{K}(\gamma_j B_j^k e_t^k) + \theta_1 P_t + \theta_2 P_t^2 + \epsilon_t$$

$$= \beta_0 + g(t) + \sum_{j=1}^{J}\gamma_j \underbrace{\sum_{k=1}^{K}(B_j^k e_t^k)}_{E_t^j} + \theta_1 P_t + \theta_2 P_t^2 + \epsilon_t \quad (7)$$

where $g(t)$ is a function that controls for trends in TFP, $E_t^j$ is the $j^{th}$ transformed temperature exposure variable and $P_t$ is precipitation.

We follow (37) in estimating non-linear effects of temperature exposure to various temperature bins (see data section). This procedure is explained in greater detail in (47). In essence, we want to estimate the effect on TFP of spending additional time in each bin. However, the exposure bins are numerous and collinear, so estimating their effects separately in the same regression is infeasible. We thus reduce the dimensionality of these bins by fitting a natural cubic spline. The underlying assumption is that the marginal effects of neighboring bins are similar and vary smoothly.

More precisely, the term $e_t^k$ represents the time spent in the $k^{th}$ temperature bin or interval. We originally compute monthly bins in 1°C intervals from -30 to 50°C. To avoid having little to no exposure in the extreme tails, we top and bottom code bins so that the extreme bins have no less than 0.1% of the total exposure. $K$ is the total number of bins after top and bottom coding the tails. The term $B_j^k$ represents the value of the $k^{th}$ row and $j^{th}$ column of the basis matrix of natural cubic spline with $J$ degrees of freedom ($J = 5$ for our baseline model). Thus $E_t^j = B_j^k e_t^k$ is a transformed temperature exposure variable. Note that we only estimate $J = 5$ parameters $\gamma_j$ in our baseline model, which greatly reduces the dimensionality of the problem.

The trend $g(t)$ can be modelled in various ways. Our baseline approach relies on a Hamilton trend with 1 year look-ahead (34). That is, $g(t)$ includes the 4 more recent lags of TFP as additional predictors. Fig. S20 illustrates the flexibility of the Hamilton trend with various look-ahead periods (panels A and C). To capture the covariance between the model in equations (6) and (7) we make sure that the bootstraps sample the sample block of observations.

We could, of course, simply set $g(t) = \beta_1 \log S_t$ as in equation (6). However, we find that this approach leads to imprecise estimates of weather effects (Fig. S7) relative to the baseline approach with a Hamilton trend (Fig. 2). We also find that control for a linear or quadratic time trend also leads to more imprecise estimates of weather effects (Fig. S10 and S11). These time trends do not appear to control well for small deviations in TFP trends over the sample period (Fig. S20, panels B and D).

Our baseline weather variables in equation (7) are aggregated over the calendar year (January–December). We consider alternative seasons as well. We apply the approach proposed and validated in (*41*) to recover the underlying seasonality of the TFP data based on various combinations of weather variables (Fig. S8). We find that seasons that best improve TFP predictions out of sample tend to be concentrated in the summer. However, we find that relying on weather variables tailored to the best fitting season does not substantially alter our findings (compare Fig. S9 and 2).

We report the sensitivities to uniform changes in temperature and precipitation in Fig. 2 (panels C and D) and in Tables S5 and S6.

*Computing impacts of climate change on TFP*

We compute estimates of the impact of climate change on TFP at year *t* using the empirical version of equation (4):

$$\widehat{D}_t = \sum_{j=1}^{J} \hat{\gamma}_j \left(E_t^j - \bar{E}_0^j\right) + \hat{\theta}_1 (P_t - \bar{P}_0) + \hat{\theta}_2 (P_t^2 - \bar{P}_0^2) \tag{8}$$

This equation combines the econometric uncertainty reflected in the $\hat{\gamma}$s and $\hat{\theta}$s, and the uncertainty stemming from the GCMs in CMIP6 reflected in $E_t^j - \bar{E}_0^j$ and $P_t - \bar{P}_0$. To combine these uncertainties, we randomly match (2000 times) bootstrapped estimates of $\hat{\gamma}$s and $\hat{\theta}$s with GCMs for each SSP scenario. In all cases, the reference climatology is computed over 1950-1960 to accommodate the sample data limitations of the CMIP6 data. The top panels of Fig. 3 show impacts that combine both econometric and climate uncertainty. The bottom panels of the figure show impact estimates separately for each GCM. We report impacts of climate change on TFP by 2025, 2050, 2075 and 2100 in Tables S7–S10.

*Solving for offsetting R&D expenditures*

We first solve for the relative change in the knowledge stock $\Delta S_t$ needed to offset the emerging impacts of climate change on TFP, $\Delta S_t$. This is directly based on the empirical version of equation (5) that incorporates an additional source of econometric uncertainty through $\hat{\beta}_1$:

$$\Delta \log \hat{S}_t^* = -\frac{\hat{D}_t}{\hat{\beta}_1} \qquad (9)$$

The term $\hat{D}_t$ is in log points, and so is $\Delta \log \hat{S}_t^*$. These estimates are depicted in the first row of panels of Fig. S17 as percentage changes. Because the research stock moves slowly, these stock changes are smoothed over time using a natural cubic spline with 3 degrees of freedom over 1950–2100. These estimates are also reported in Table S11 (top).

We also convert these relative changes in the knowledge stock to monetary values $\Delta \hat{S}_t^*$. This requires we need to assume a counterfactual level of the knowledge stock without climate change. For this, we assume that public R&D spending $RD_t$ remains unchanged (in real terms) at its 2020 level over 2021–2100. We then compute the associated counterfactual knowledge stock over 1950–2100. We can then derive a monetary value of $\Delta \hat{S}_t^*$ by subtracting the level of the knowledge stock in the scenario without climate change from the knowledge stock in the scenario with climate change $S_t^*$. These estimates are show in the second row of Fig. S17. These estimates are also reported in Table S11 (bottom). These estimates can be interpreted as the additional change in the knowledge stock needed to avoid declines in TFP relative to a counterfactual scenario without climate change *and* stagnant TFP. If one assumes TFP grows in the counterfactual without climate change (due to rising R&D spending) this would lead to even larger monetary values of $S_t^*$. Thus, our monetary values of these estimates can be interpreted as a lower bound. In addition, note that a $1B increase in the knowledge stock in a steady state, requires a *permanent* increase of $1B/year in public R&D spending.

We then seek to solve for spending trajectories $RD_t$ from 2021 to various target years $T$ that would be necessary to achieve $\Delta \hat{S}_t^*$. That is, to achieve a change in the knowledge stock that fully compensates for the impacts of climate change by year $T$. There is, however, an infinite number of annual spending pathways to achieve a certain change in the knowledge stock. For simplicity, we consider spending pathways with fixed annual growth rates which allows direct comparison with historical spending trajectories.

Based on our estimates $\hat{\beta}_1$, we can simulate the growth of $S_t$ under a wide range of fixed annual spending growth rates from 2021 until a target year $T$. We can then contrast the attained changes in the knowledge stock in year $T$, $\Delta S_T$, with our estimates of the climate-offsetting changes in the knowledge stock $\Delta \hat{S}_T^*$. We then select the fixed annual growth rates that approximate the necessary changes in the knowledge stock. Noting that $\Delta \hat{S}_T^*$ is an estimate, we derive the necessary spending growth for the central estimates and for its 95th and 99th confidence interval. This mapping between fixed annual growth rates and offsetting changes in the knowledge stock is depicted in Fig. S18 for each SSP scenario. Naturally, the more imminent the target date $T$ is, the more rapid the growth in annual spending $RD_t$ needs to be to reach a certain change in the knowledge stock $\Delta \hat{S}_t^*$. The third row of panels in Fig. S17 shows the necessary changes in annual public R&D spending to offset climate change impacts for various target years from 2050 to 2100 (note these are also represented in the top left quadrants of panels in Fig. S18). These estimates are also reported in Table S12.

We also track the additional cumulative spending necessary to offset climate change impacts by various target years. These are depicted in the fourth row of panels of Fig. S17. These estimates are also reported in Table S13.

# List of SM Figures and Tables

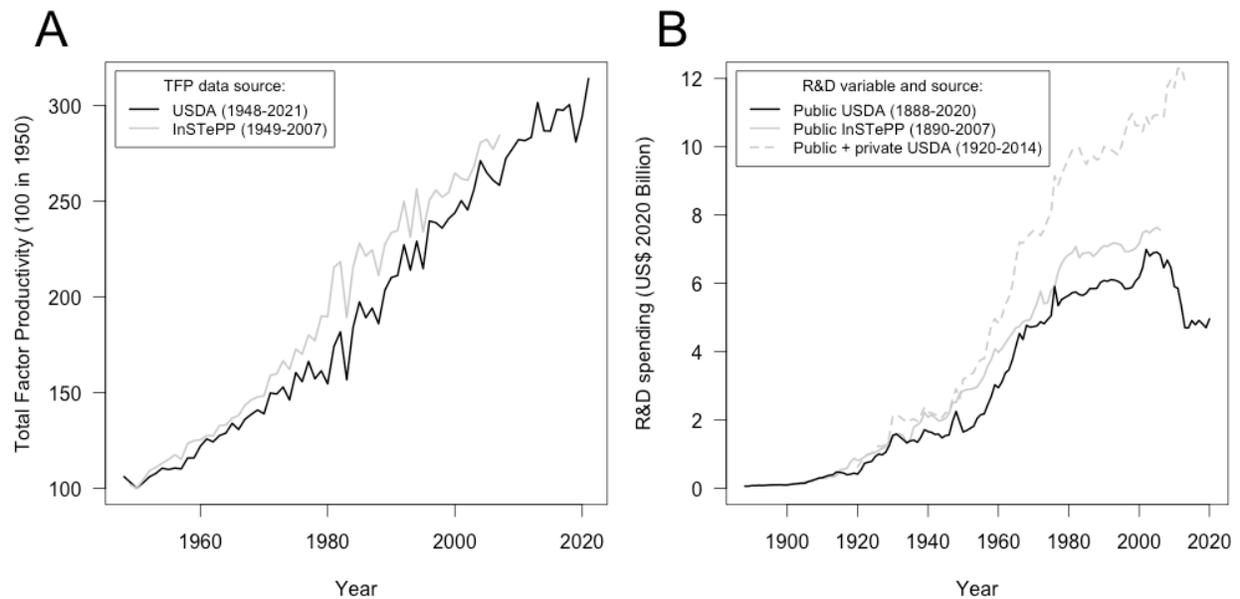

**Fig. S1.**
**Agricultural Total Factor Productivity (TFP) and Research & Development (R&D) expenditures.** Panel A shows the official agricultural TFP data from the US Department of Agriculture (USDA)'s Economic Research Service (ERS) in black. The grey line represents an alternative dataset from the International Science & Technology Practice & Policy (InSTePP) group. Panel B shows the official USDA-ERS data on annual expenditures in public agricultural R&D spending in black. The grey line represents the alternative dataset from InSTePP. The dotted grey line represents total (public and private) agricultural R&D expenditures obtained from USDA-ERS.

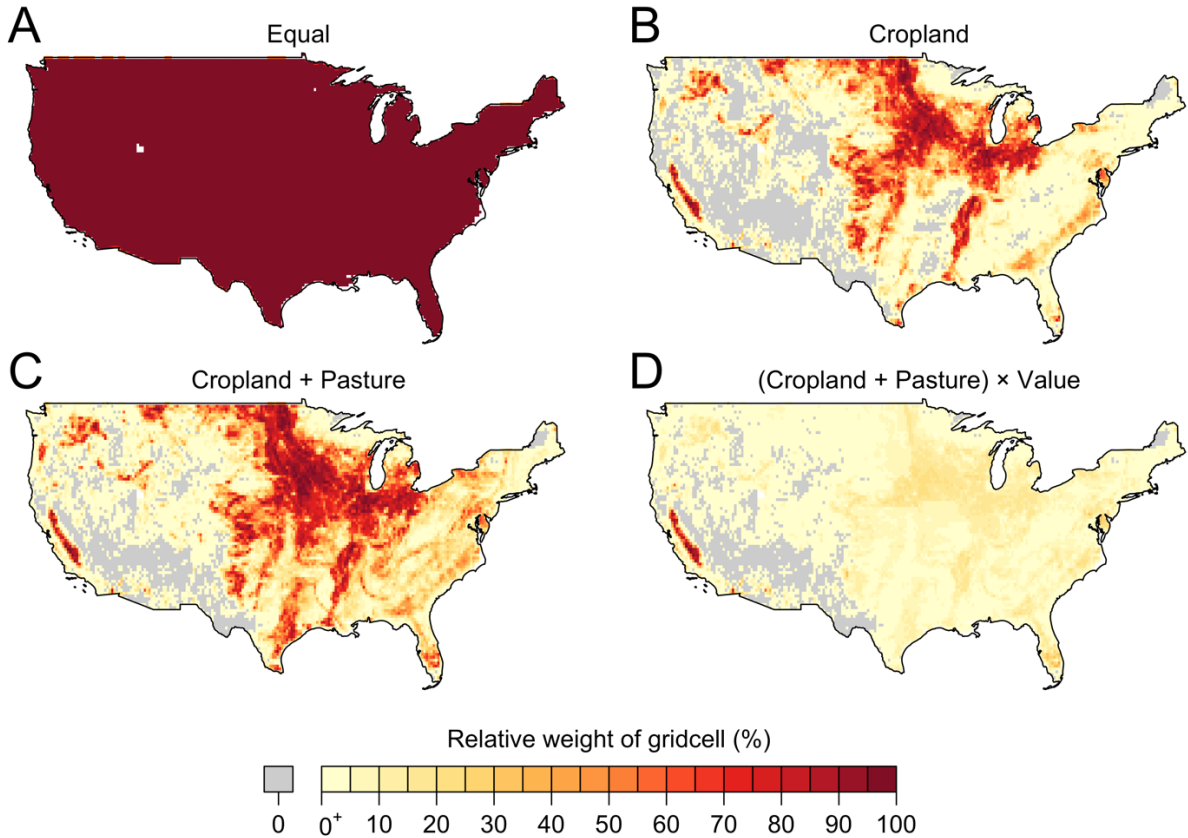

**Fig. S2.**
**Spatial weights to aggregate weather data to the national level.** Panel A shows basic equal weights. Panel B shows weights based on the proportion of weather grid cells occupied by cropland. Panel C is similar but incorporates both cropland and pastures. Panel D scales weights in panel C by state based on their overall share of national value produced. These corresponds to the baseline weights we use in the paper. The grid cells shown here correspond to the ERA-5 dataset. For each panel, weights were scaled from 0 to 100 to show relative importance. The actual weights used for aggregation sum to 1.

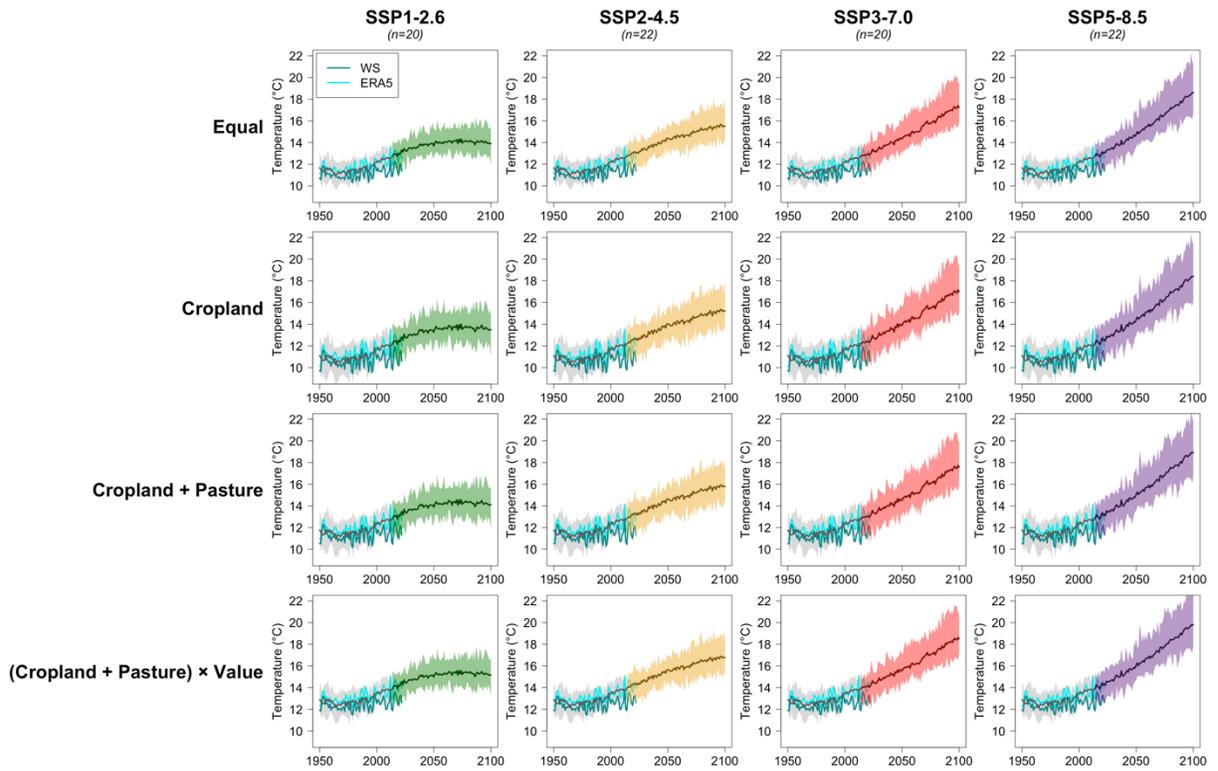

**Fig. S3.**
**CMIP6 projections for mean annual temperature.** Each column of panel corresponds to a Shared Socioeconomic Pathway (SSP) scenario under CMIP6. The number in parenthesis under the column title indicates the number of GCMs represented in the corresponding column. Grey lines and bands correspond to the historical ("hist") experiments (1950–2014). The SSPs are represented for the rest of the century (2015–2100). For reference, we show observed historical values for ERA-5 and for the WS dataset. Each row of panels shows the climate data aggregated according to the various aggregation weights show in on Fig. S2.

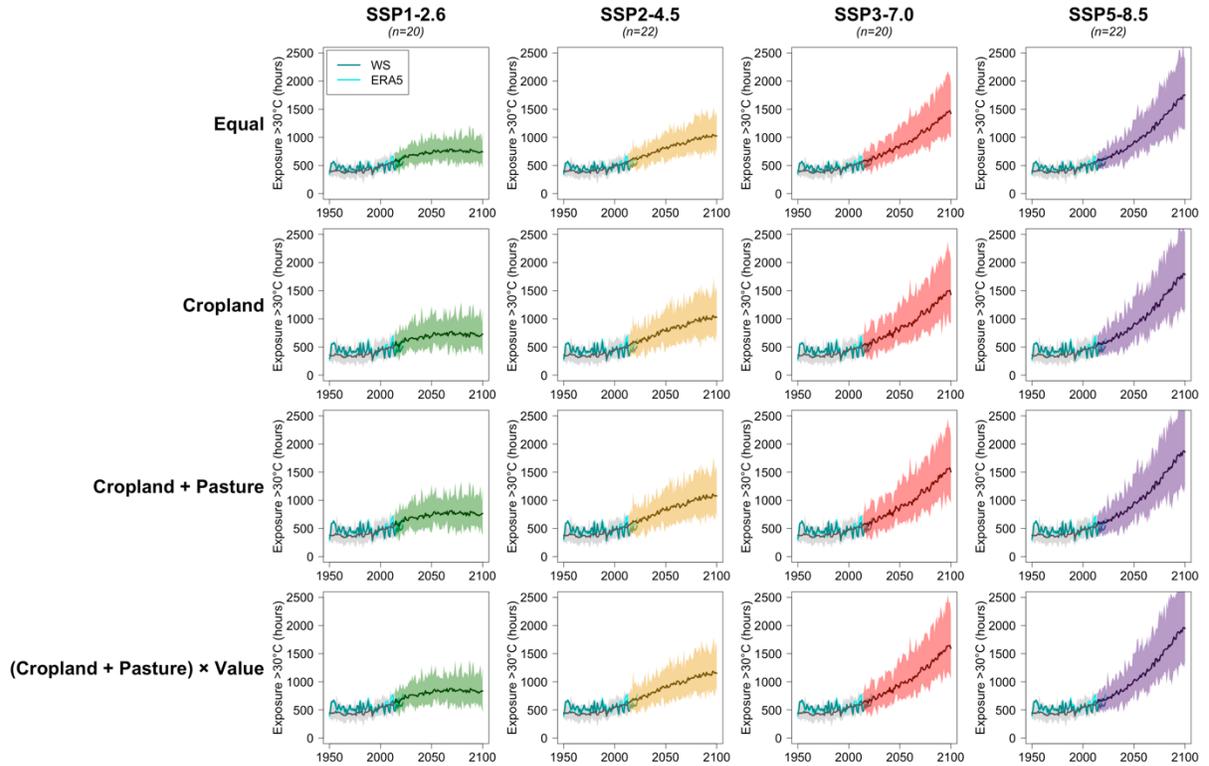

**Fig. S4.**
**CMIP6 projections for annual exposure above 30°C.** Each column of panel corresponds to a Shared Socioeconomic Pathway (SSP) scenario under CMIP6. The number in parenthesis under the column title indicates the number of GCMs represented in the corresponding column. Grey lines and bands correspond to the historical ("hist") experiments (1950–2014). The SSPs are represented for the rest of the century (2015–2100). For reference, we show observed historical values for ERA-5 and for the WS dataset. Each row of panels shows the climate data aggregated according to the various aggregation weights show in on Fig. S2.

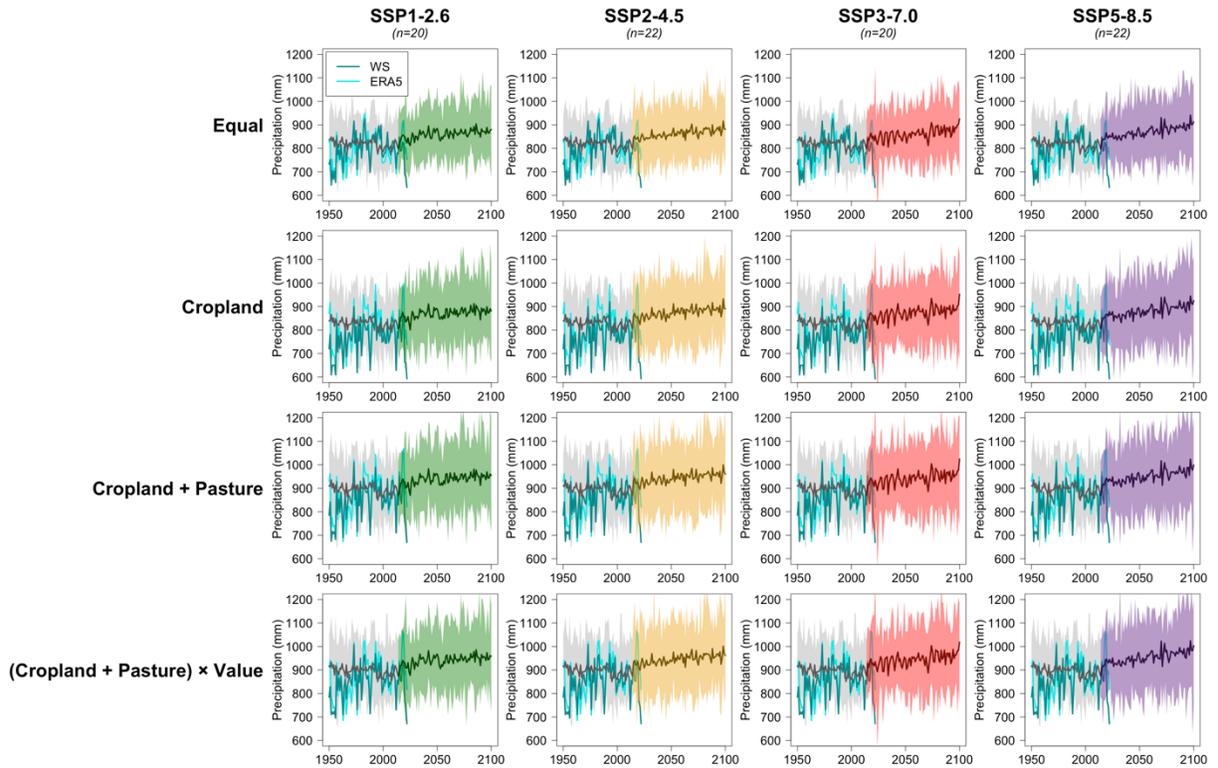

**Fig. S5.**

**CMIP6 projections for annual precipitation.** Each column of panel corresponds to a Shared Socioeconomic Pathway (SSP) scenario under CMIP6. The number in parenthesis under the column title indicates the number of GCMs represented in the corresponding column. Grey lines and bands correspond to the historical ("hist") experiments (1950–2014). The SSPs are represented for the rest of the century (2015–2100). For reference, we show observed historical values for ERA-5 and for the WS dataset. Each row of panels shows the climate data aggregated according to the various aggregation weights show in on Fig. S2.

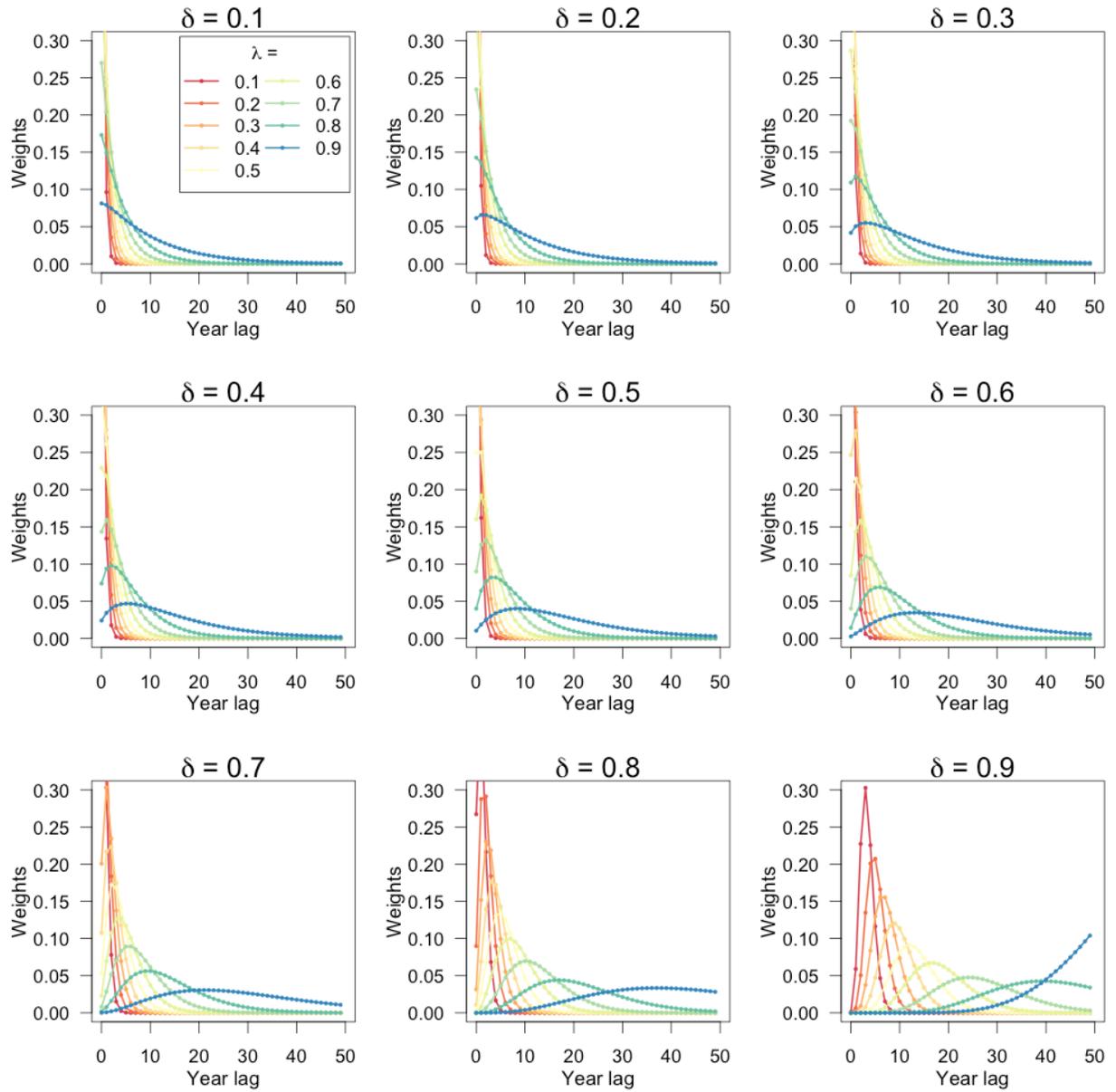

**Fig. S6.**
**Lag weights based on the gamma distribution for constructing the knowledge stock.** We conduct our grid search over values ranging from 0.05 and 0.95 at 0.05 intervals to find the optimal values of the $\lambda$ and $\delta$ parameters.

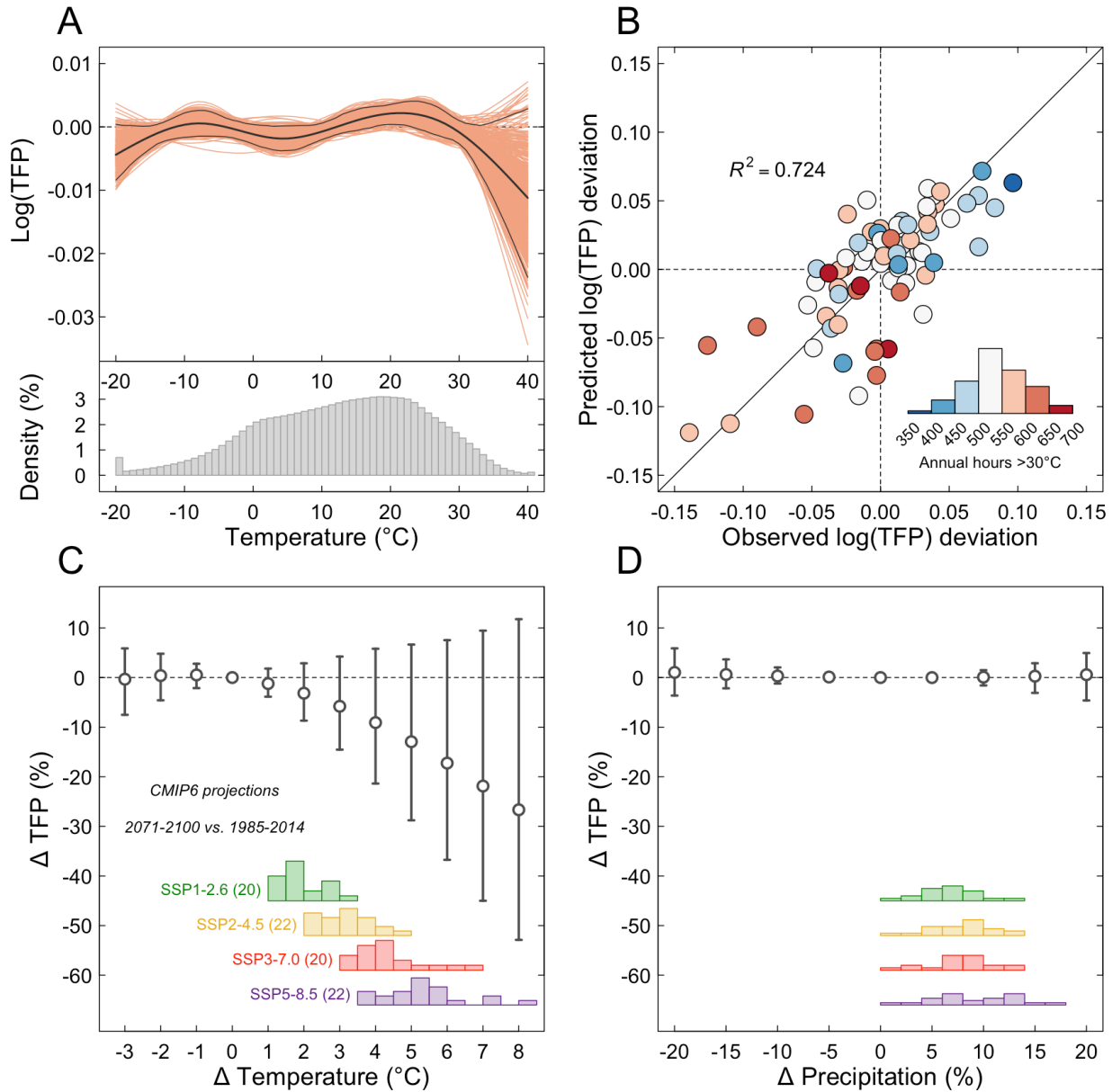

**Fig. S7.**
**Historical impact of weather fluctuations on US agricultural TFP based on model with only knowledge stock controlling for the trend.** This figure is analogous to Fig. 2 in the paper but controls for trends differently. (**A**) Effect of additional annual exposure to various levels of temperature on TFP. The various red lines represent 500 bootstrapped estimates. Black lines represent 95% pointwise confidence intervals. (**B**) Scatterplot of observed and predicted TFP deviations from a linear trend. (**C**) Impact of uniform changes in annual temperature on TFP. (**D**) Impact of uniform changes in annual precipitation on TFP. For reference, panels C and D show the distribution of changes in annual temperature and precipitation from GCMs in CMIP6 between the end of the century (2071–2100) and a historical reference period (1985–2014).

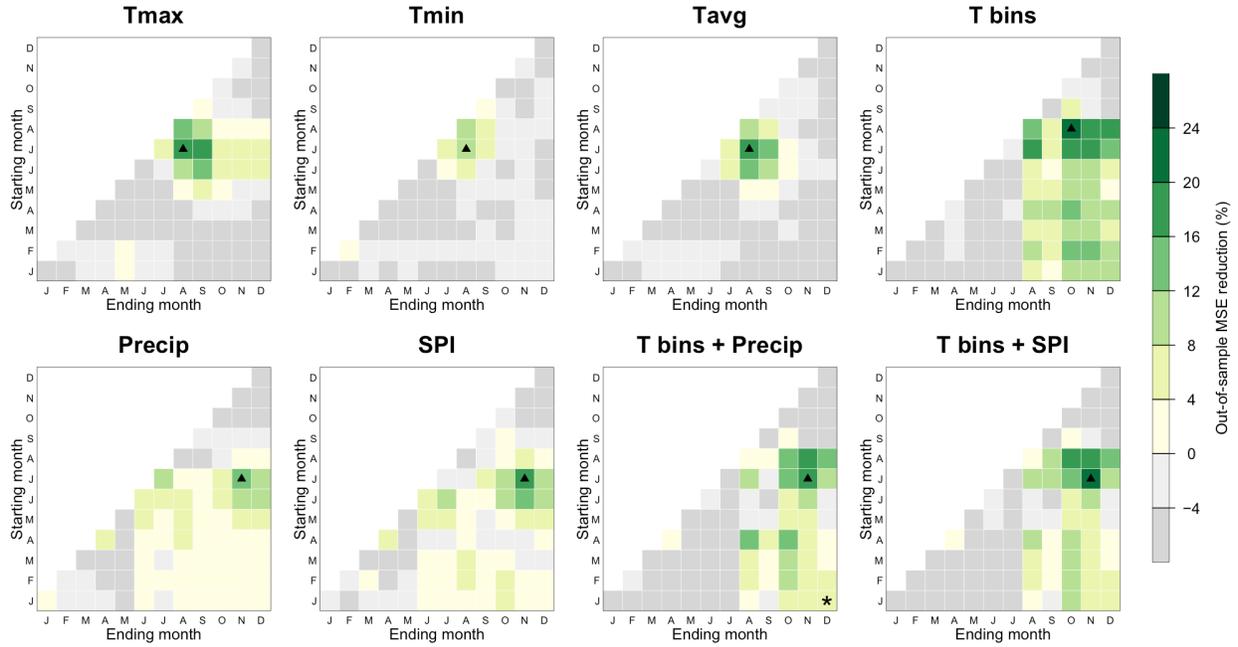

**Fig. S8.**
**Reduction in out-of-sample MSE for TFP models based on alternative weather variables and seasons.** Each panel corresponds to models using a specific set of weather variables as predictors of TFP. Grid cells within each panel show the MSE reductions of a specific season during a leave-one-year-out cross-validation relative to a model without weather variables. All models control for trends in TFP using a Hamilton filter. Green cells indicate weather variables improve prediction of TFP out of sample. The best fitting season is highlighted with a black triangle. Our baseline model, marked with a star, is based on the "T bins + Precip" model using the calendar year (January–December) as the relevant season.

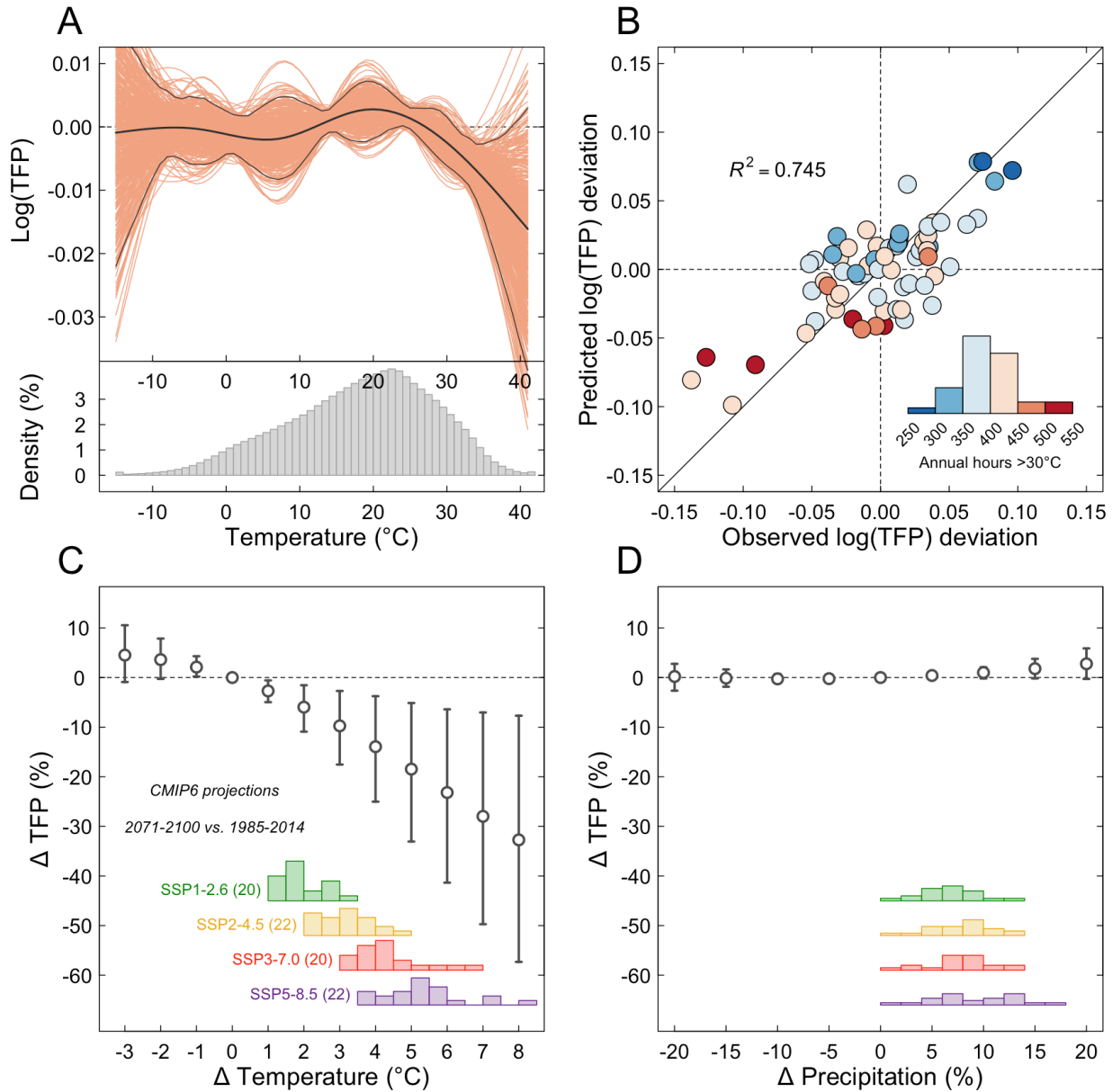

**Fig. S9.**
**Historical impact of weather fluctuations on US agricultural TFP based on weather variables aggregated over the best-fitting season.** This figure is analogous to Fig. 2 in the paper but relies on different seasons for the aggregation of weather variables. (**A**) Effect of additional annual exposure to various levels of temperature on TFP. The various red lines represent 500 bootstrapped estimates. Black lines represent 95% pointwise confidence intervals. (**B**) Scatterplot of observed and predicted TFP deviations from a linear trend. (**C**) Impact of uniform changes in annual temperature on TFP. (**D**) Impact of uniform changes in annual precipitation on TFP. For reference, panels C and D show the distribution of changes in annual temperature and precipitation from GCMs in CMIP6 between the end of the century (2071–2100) and a historical reference period (1985–2014).

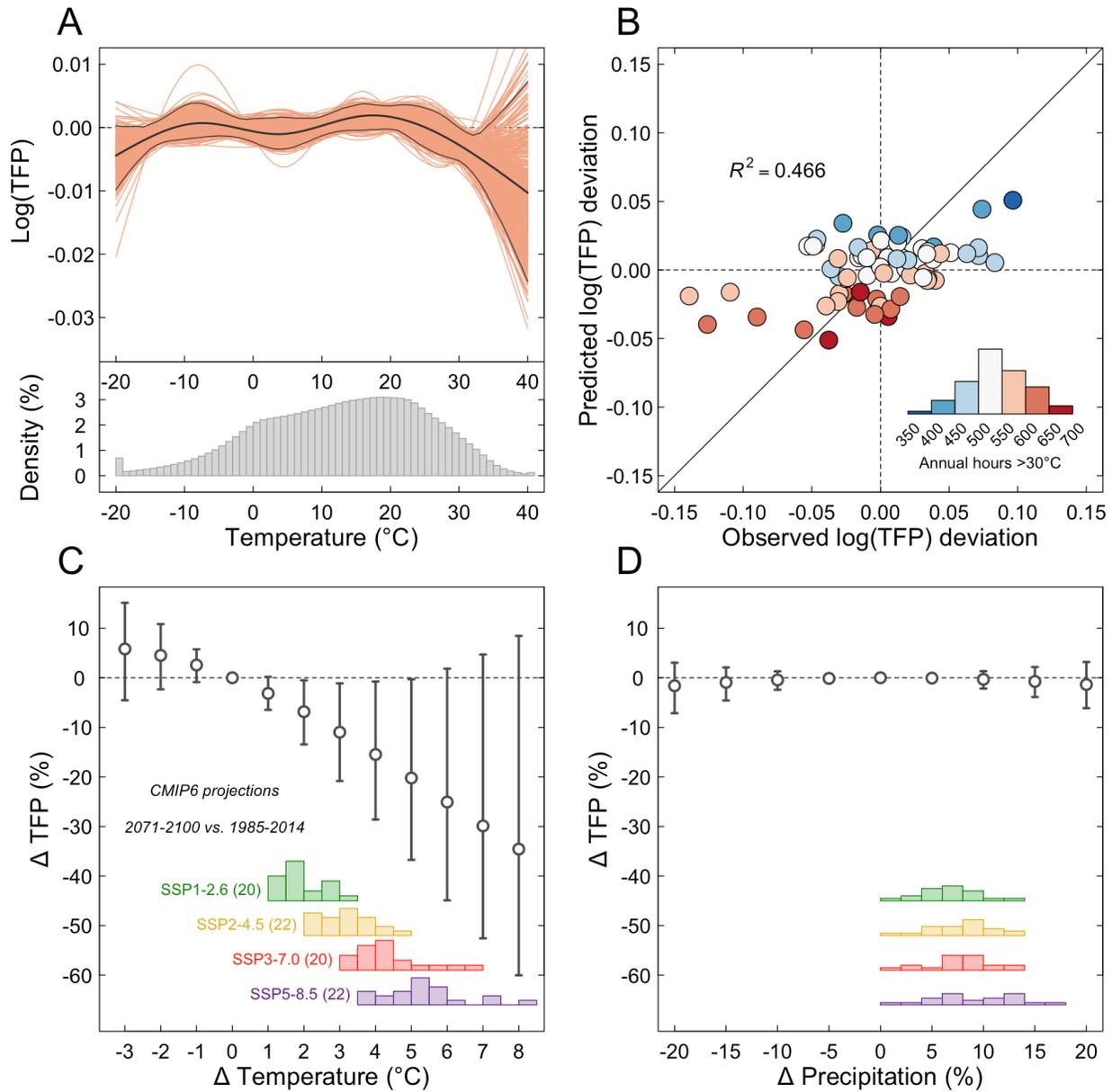

**Fig. S10.**

**Historical impact of weather fluctuations on US agricultural TFP based on model controlling for a linear time trend.** This figure is analogous to Fig. 2 in the paper but relies on a linear time trend instead of a Hamilton filter. (**A**) Effect of additional annual exposure to various levels of temperature on TFP. The various red lines represent 500 bootstrapped estimates. Black lines represent 95% pointwise confidence intervals. (**B**) Scatterplot of observed and predicted TFP deviations from a linear trend. (**C**) Impact of uniform changes in annual temperature on TFP. (**D**) Impact of uniform changes in annual precipitation on TFP. For reference, panels C and D show the distribution of changes in annual temperature and precipitation from GCMs in CMIP6 between the end of the century (2071–2100) and a historical reference period (1985–2014).

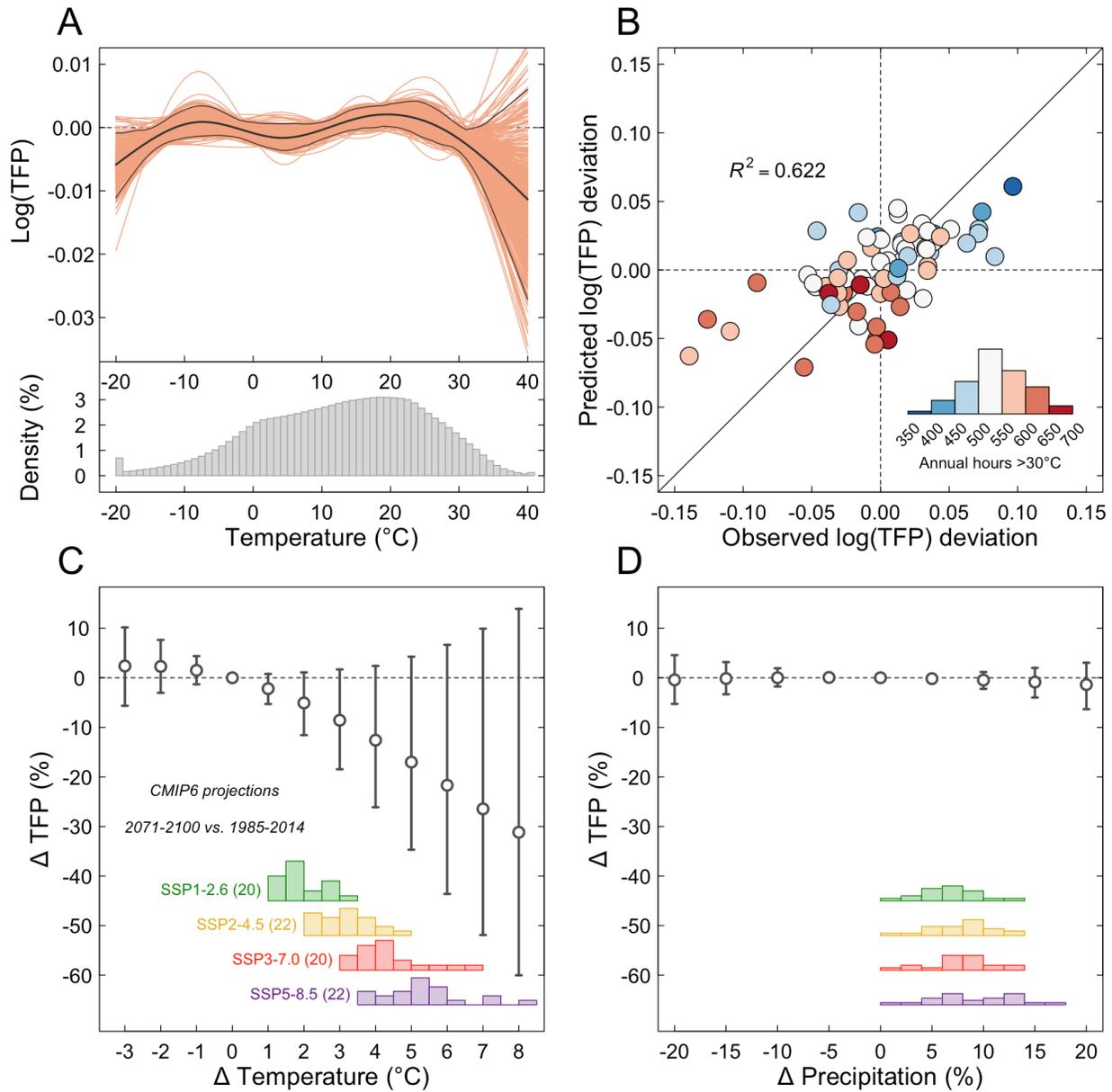

**Fig. S11.**

**Historical impact of weather fluctuations on US agricultural TFP based on model controlling for a quadratic time trend.** This figure is analogous to Fig. 2 in the paper but relies on a quadratic time trend instead of a Hamilton filter. (**A**) Effect of additional annual exposure to various levels of temperature on TFP. The various red lines represent 500 bootstrapped estimates. Black lines represent 95% pointwise confidence intervals. (**B**) Scatterplot of observed and predicted TFP deviations from a linear trend. (**C**) Impact of uniform changes in annual temperature on TFP. (**D**) Impact of uniform changes in annual precipitation on TFP. For reference, panels C and D show the distribution of changes in annual temperature and precipitation from GCMs in CMIP6 between the end of the century (2071–2100) and a historical reference period (1985–2014).

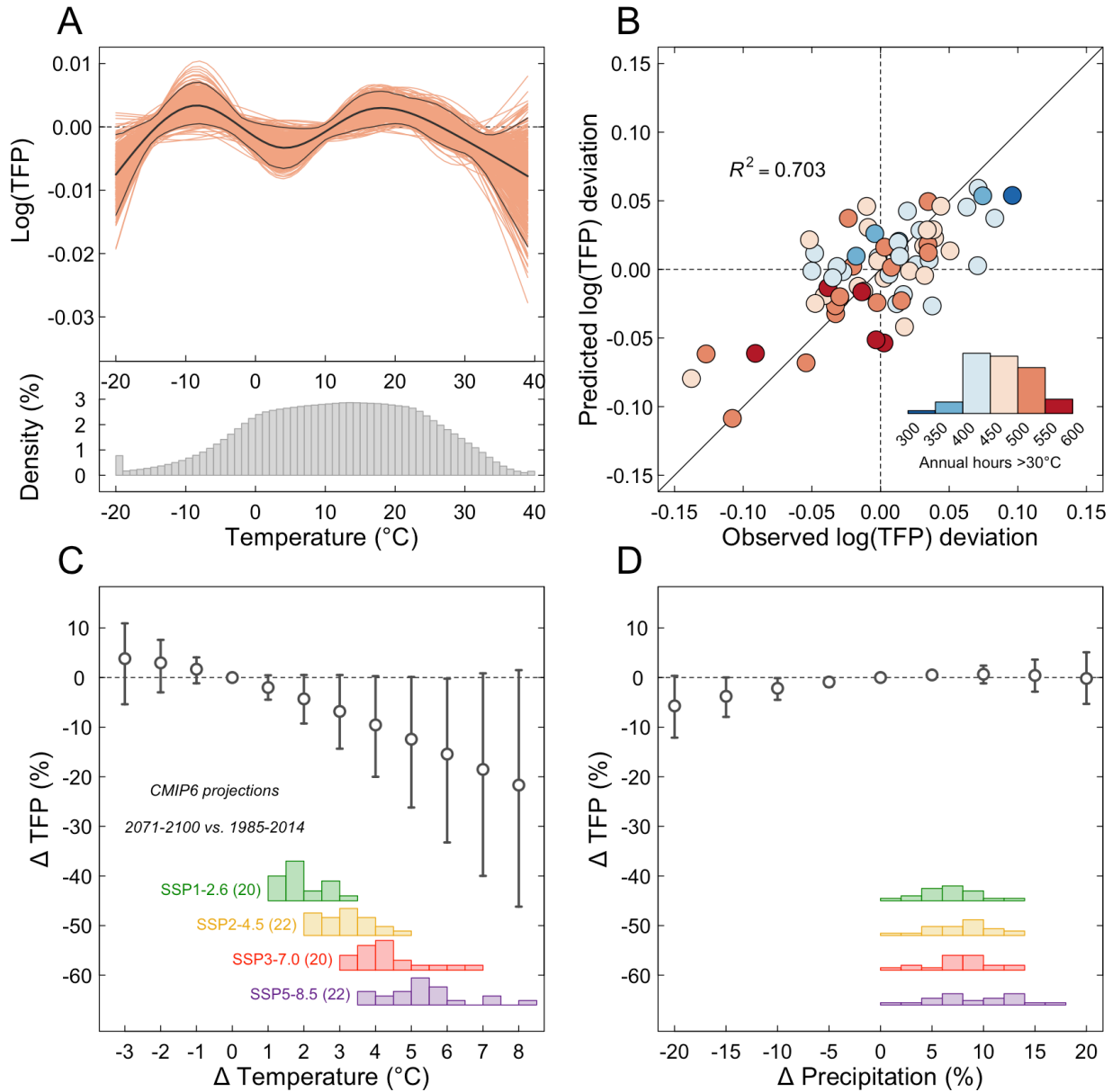

**S12.**

**Historical impact of weather fluctuations on US agricultural TFP based on weather variables aggregated using equal spatial weights.** This figure is analogous to Fig. 2 in the paper but relies on weather data aggregated using equal spatial weights (see Fig. S2). (**A**) Effect of additional annual exposure to various levels of temperature on TFP. The various red lines represent 500 bootstrapped estimates. Black lines represent 95% pointwise confidence intervals. (**B**) Scatterplot of observed and predicted TFP deviations from a linear trend. (**C**) Impact of uniform changes in annual temperature on TFP. (**D**) Impact of uniform changes in annual precipitation on TFP. For reference, panels C and D show the distribution of changes in annual temperature and precipitation from GCMs in CMIP6 between the end of the century (2071–2100) and a historical reference period (1985–2014).

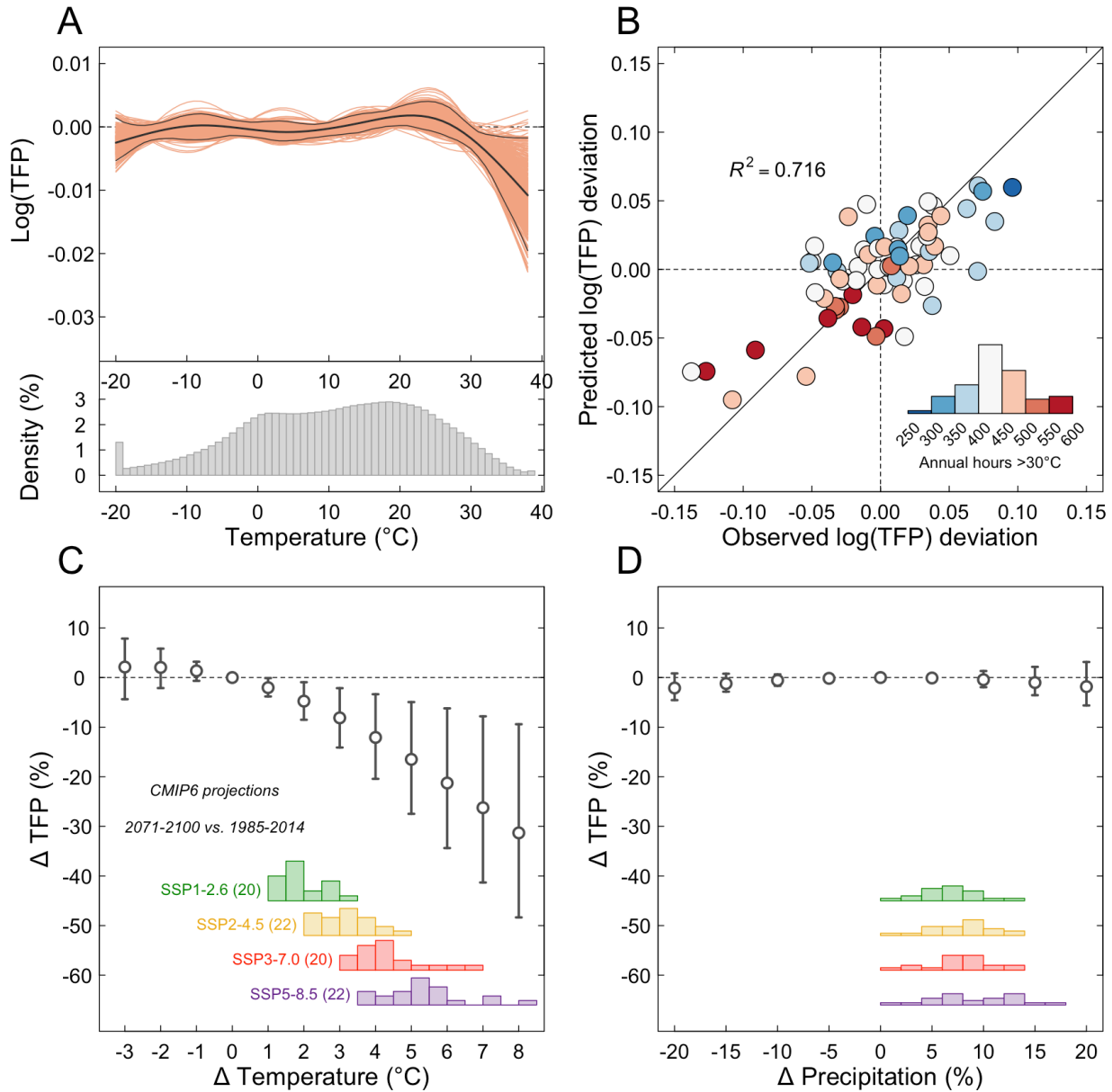

**S13.**

**Historical impact of weather fluctuations on US agricultural TFP based on weather variables aggregated using cropland spatial weights.** This figure is analogous to Fig. 2 in the paper but relies on weather data aggregated using cropland spatial weights (see Fig. S2). (**A**) Effect of additional annual exposure to various levels of temperature on TFP. The various red lines represent 500 bootstrapped estimates. Black lines represent 95% pointwise confidence intervals. (**B**) Scatterplot of observed and predicted TFP deviations from a linear trend. (**C**) Impact of uniform changes in annual temperature on TFP. (**D**) Impact of uniform changes in annual precipitation on TFP. For reference, panels C and D show the distribution of changes in annual temperature and precipitation from GCMs in CMIP6 between the end of the century (2071–2100) and a historical reference period (1985–2014).

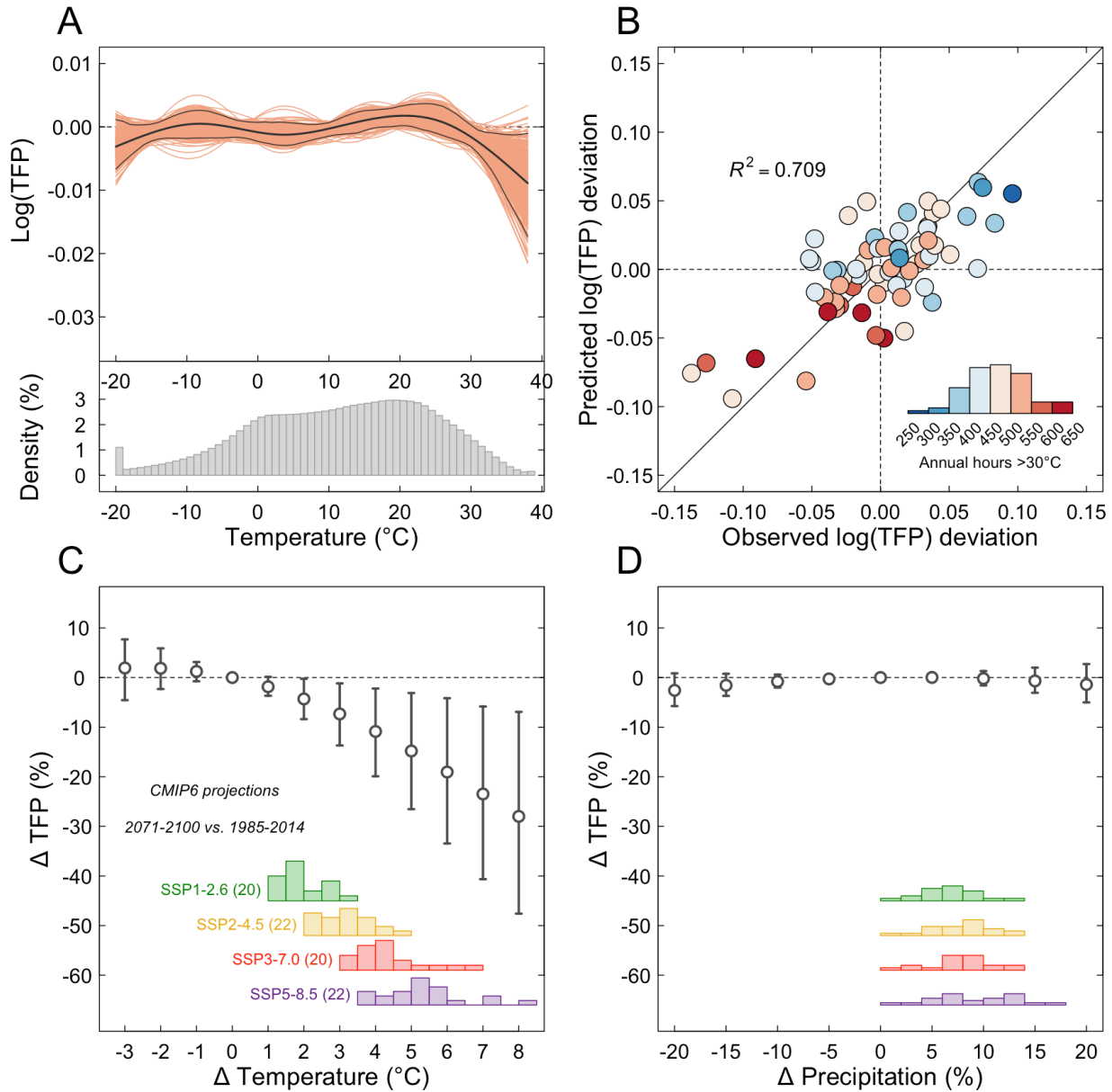

**S14.**

**Historical impact of weather fluctuations on US agricultural TFP based on weather variables aggregated using cropland and pasture spatial weights.** This figure is analogous to Fig. 2 in the paper but relies on weather data aggregated using cropland and pasture spatial weights (see Fig. S2). (**A**) Effect of additional annual exposure to various levels of temperature on TFP. The various red lines represent 500 bootstrapped estimates. Black lines represent 95% pointwise confidence intervals. (**B**) Scatterplot of observed and predicted TFP deviations from a linear trend. (**C**) Impact of uniform changes in annual temperature on TFP. (**D**) Impact of uniform changes in annual precipitation on TFP. For reference, panels C and D show the distribution of changes in annual temperature and precipitation from GCMs in CMIP6 between the end of the century (2071–2100) and a historical reference period (1985–2014).

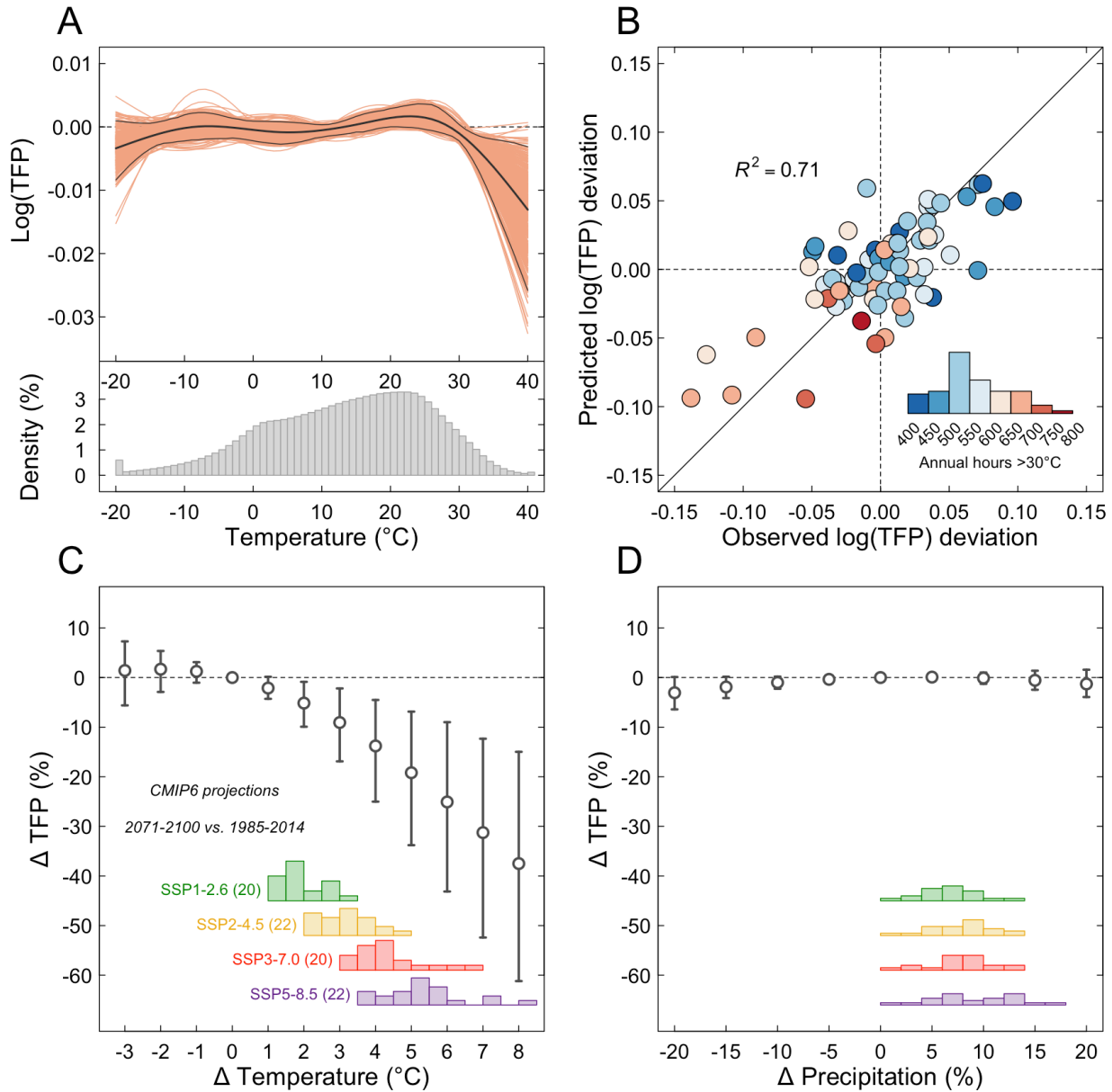

**S15.**

**Historical impact of weather fluctuations on US agricultural TFP based on the ERA5 weather data.** This figure is analogous to Fig. 2 in the paper but relies on the ERA5 weather data. (**A**) Effect of additional annual exposure to various levels of temperature on TFP. The various red lines represent 500 bootstrapped estimates. Black lines represent 95% pointwise confidence intervals. (**B**) Scatterplot of observed and predicted TFP deviations from a linear trend. (**C**) Impact of uniform changes in annual temperature on TFP. (**D**) Impact of uniform changes in annual precipitation on TFP. For reference, panels C and D show the distribution of changes in annual temperature and precipitation from GCMs in CMIP6 between the end of the century (2071–2100) and a historical reference period (1985–2014).

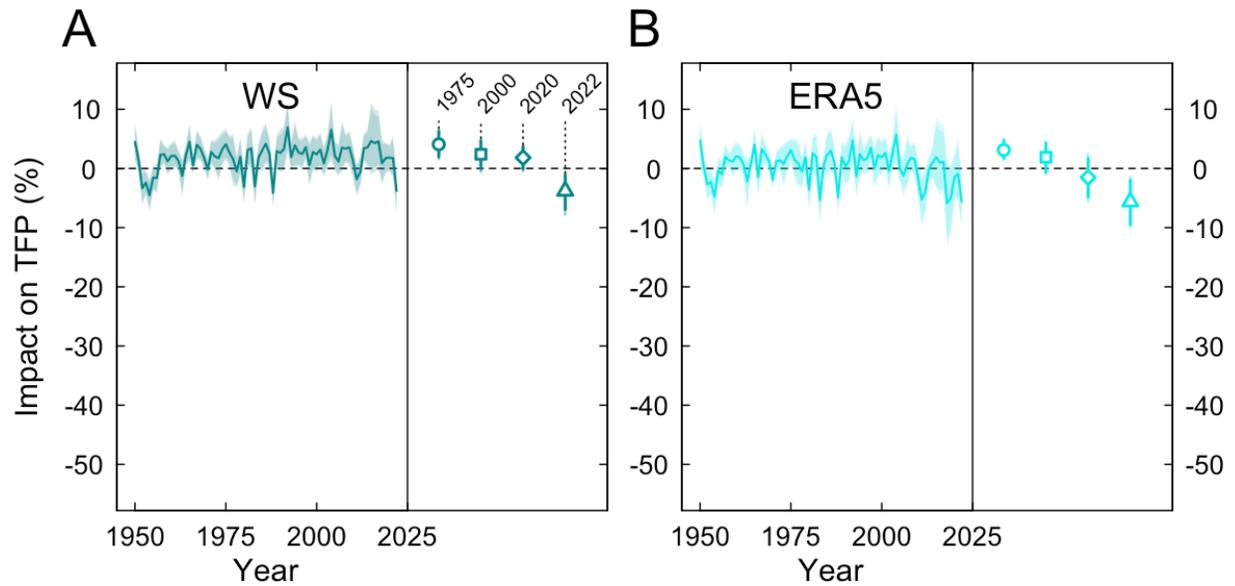

**S16.**

**Impact of observed historical weather trends on US agricultural TFP.** This figure is analogous to Fig. 3 in the main text but is based on observed historical weather data over 1950–2022. Both panels rely on 1950–1960 as a reference period. These estimates reflect both the econometric uncertainty reflects in the estimated parameters (related to $\hat{\beta}_2$) and the weather trends observed for alternative weather datasets. Note that both panels rely on parameters estimated using the WS weather data. (**A**) Impact of weather trends in WS dataset. (**B**) Impact of weather trends in ERA5 dataset.

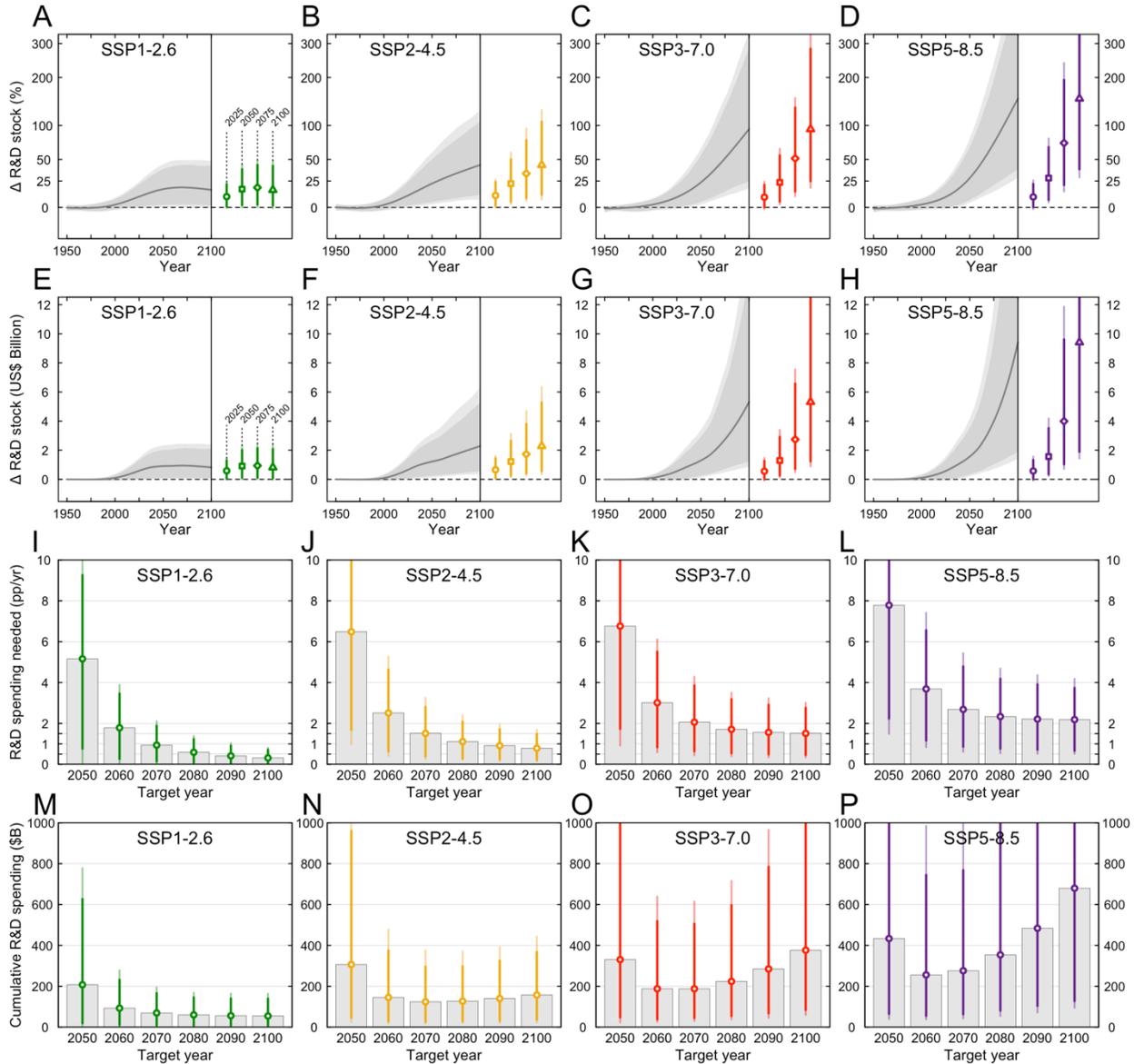

**S17.**

**R&D needs to offset climate change effects on US agricultural TFP.** All estimates show 95 (dark) and 99 percent (light) confidence intervals. (**A–D**) The relative change in the knowledge stock $\Delta S_t$ (in %) needed to offset emerging TFP impacts of climate change based on CMIP6 shown in Fig. 3. We also report these estimates on the upper part of Table S11. (**E–H**) The absolute change in the knowledge stock $\Delta S_t$ (in US& billions) needed to offset emerging TFP impacts of climate change. These estimates assume a counterfactual level $S_t$ without climate change that is constant (at its 2020 level) through the rest of the century. We also report these estimates on the bottom of Table S11. (**I–L**) The annual public R&D spending growth needed to offset climate change impacts on agricultural TFP for several target dates. (**M–P**) The cumulative public R&D spending to offset climate change impacts on agricultural TFP for several target dates. We also report these estimates on Table S12 and S13.

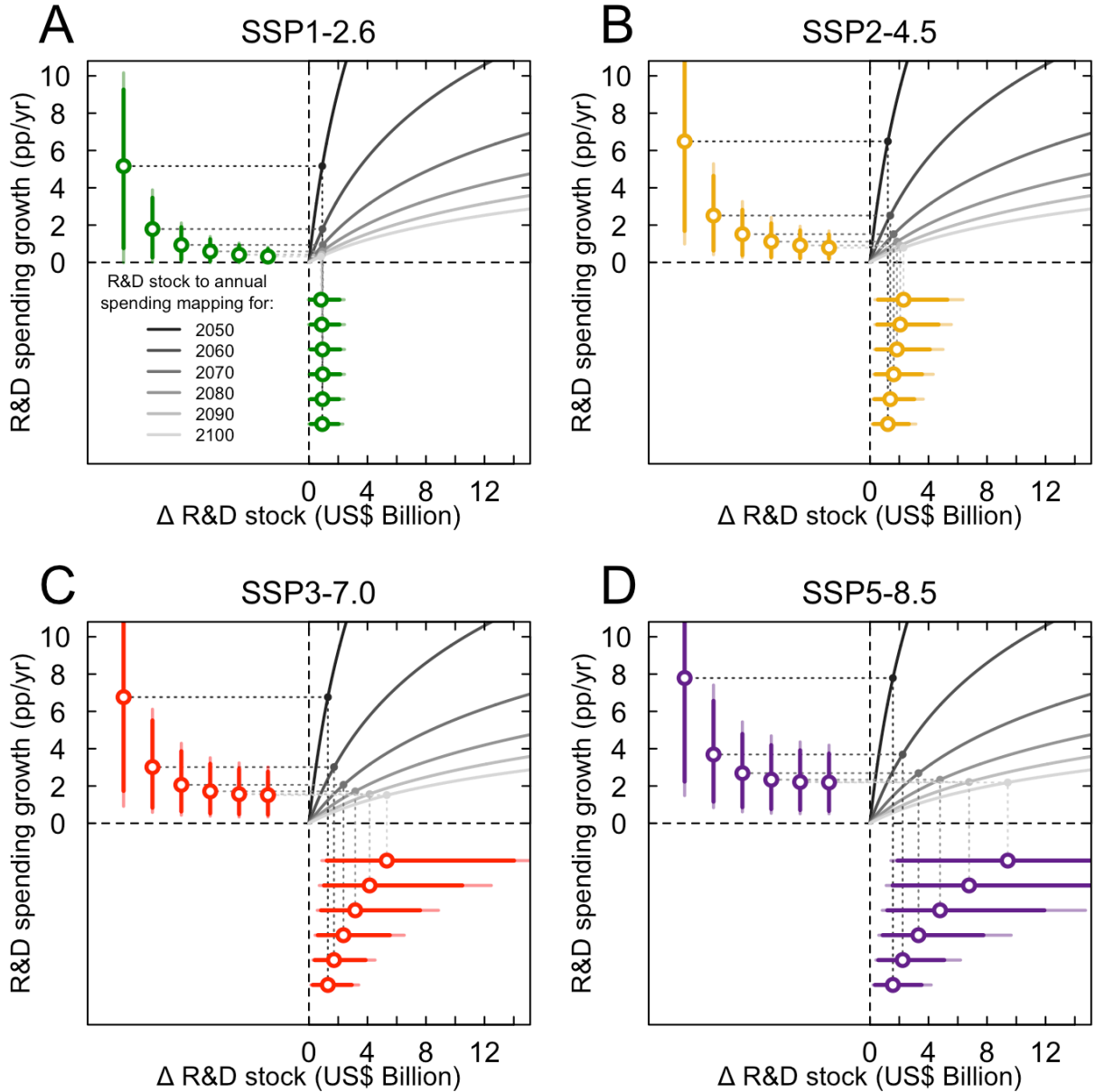

**S18.**

**Mapping changes in R&D spending to changes in the knowledge stock.** Each panel shows 6 functions colored in grey and corresponding to different target years (ranging from 2050 to 2100). These functions illustrate that spending growth needs to be larger to reach a given increase in the knowledge stock, the closer the target year is in time (2050 function is steeper than the 2100 function). Estimates shown in the bottom right quadrant of each panel reproduces estimates from Fig. S16 (second row of panels) showing the absolute change in the knowledge stock $\Delta S_t$ needed to offset climate change impacts on TFP. The corresponding spending growth necessary to obtain knowledge stock changes are shown in the upper left quadrant of each panel. All estimates show 95 (dark) and 99 percent (light) confidence intervals. Panels correspond to different SSP scenarios in CMIP6.

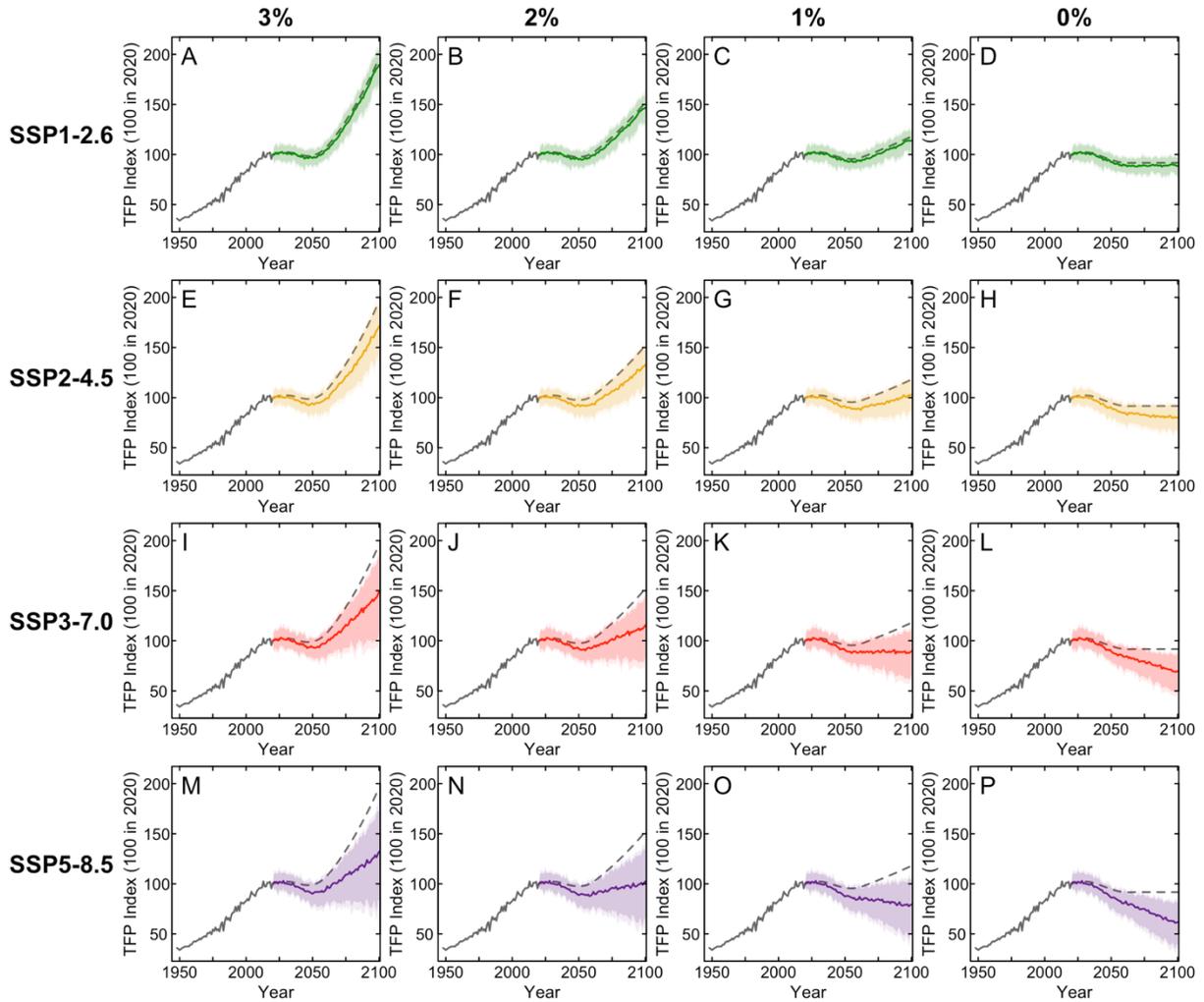

**S19.**

**Simulated TFP growth under alternative climate change and R&D spending growth scenarios.** Data for 1948–2014 corresponds to observed TFP. Each row of panels corresponds to a SSP scenario in CMIP6. Each column of panels corresponds to an annual public R&D spending growth scenario.

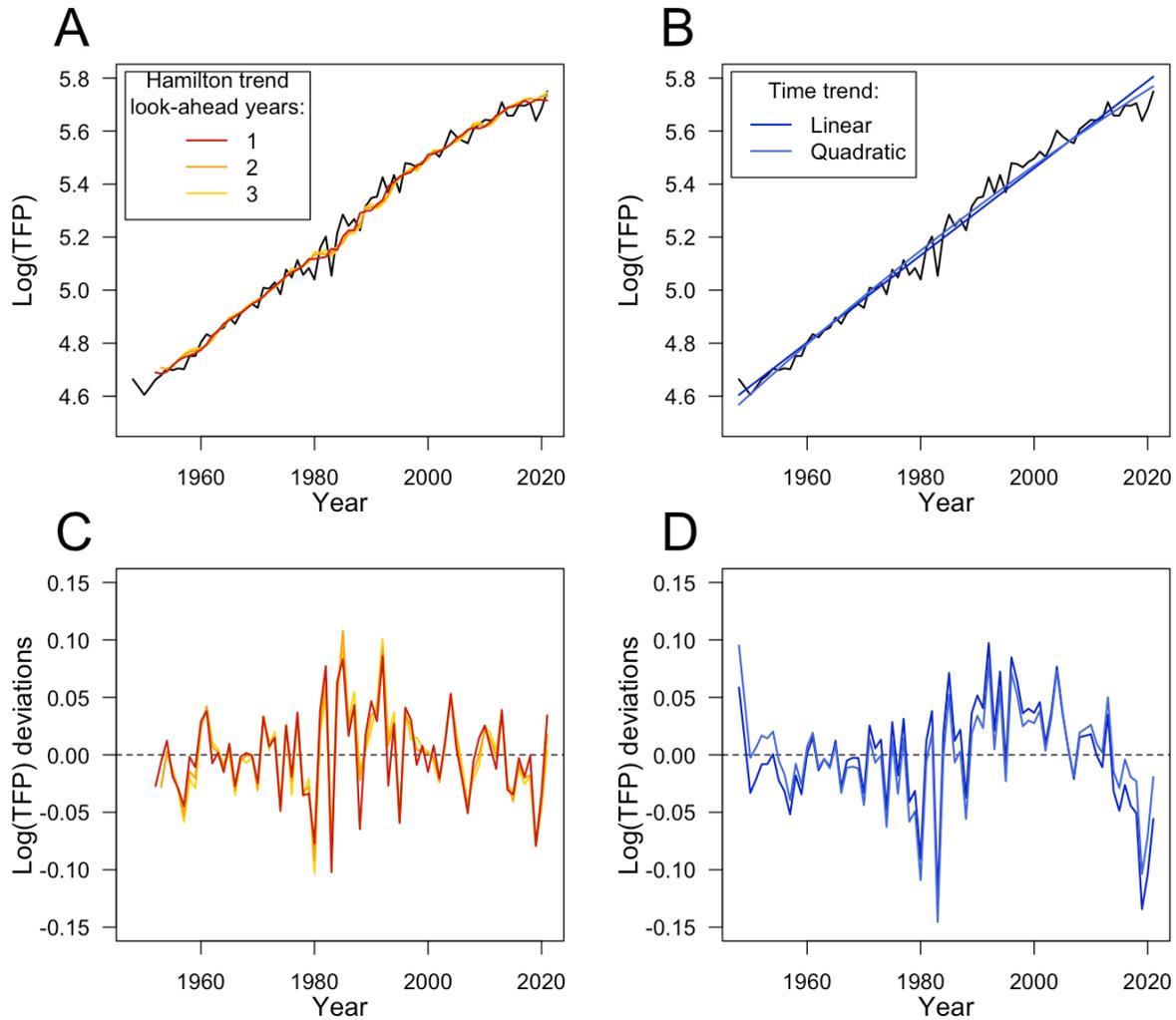

**S20.**

**Illustration of alternative controls for the TFP trend.** (**A**–**B**) The panels show the observed log TFP (in black) along with alternative trend controls including the Hamilton trend and time trends (see SM for details). (**C**–**D**) These panels show the deviations (residuals) around the predicted trend using the trend controls highlighted in the top panels, respectively. We maintain the same color scheme between top and bottom panels.

**Table S1.**
**Elasticity of the knowledge stock with alternative control variables**

| Dep. Variable: log(TFP) | (1) | (2) | (3) | (4) |
|---|---|---|---|---|
| Log of knowledge stock ($\beta_1$) | 0.503** | 0.497** | 0.335** | 0.474** |
|  | (0.013) | (0.015) | (0.117) | (0.154) |
| Controls: |  |  |  |  |
| Weather | Yes | No | Yes | Yes |
| Trend | No | No | Linear | Hamilton filter |
| Degrees of freedom | 62 | 71 | 61 | 54 |
| Observations | 71 | 73 | 71 | 67 |

Notes: These models are based on the best-fitting measure of the knowledge stock variable corresponding to research lags defined with a gamma distribution with parameters $\lambda = 0.75$ and $\delta = 0.9$ over 50 years. Standard errors are in parenthesis and are computed based on an overlapping block bootstrap to account to serial correlation (block size = 5 obs). ** $p < 0.01$, * $p < 0.05$. See SM for more details.

**Table S2.**
**Elasticity of the knowledge stock using the total (public & private) R&D expenditures**

| Dep. Variable: log($TFP$) | (1) | (2) | (3) | (4) |
|---|---|---|---|---|
| Log of knowledge stock ($\beta_1$) | 0.453** | 0.462** | 0.359 | 0.609** |
|  | (0.012) | (0.013) | (0.246) | (0.221) |
| Controls: |  |  |  |  |
| Weather | Yes | No | Yes | Yes |
| Trend | No | No | Linear | Hamilton filter |
| Degrees of freedom | 37 | 44 | 36 | 29 |
| Observations | 46 | 46 | 46 | 42 |

Notes: This is similar to Table S1 but the knowledge stock is constructed from the sum of both public and private R&D expenditures. As a result, the number of observations is smaller due to the shorter time series of the private R&D data. These models are based on the best-fitting measure of the total knowledge stock variable corresponding to research lags defined with a gamma distribution with parameters $\lambda = 0.8$ and $\delta = 0.9$ over 50 years. Standard errors are in parenthesis and are computed based on an overlapping block bootstrap to account to serial correlation (block size = 5 obs). ** $p < 0.01$, * $p < 0.05$. See SM for more details.

**Table S3.**
**Elasticity of the knowledge stock using alternative agricultural data source**

| Dep. Variable: log(TFP) | (1) | (2) | (3) | (4) |
|---|---|---|---|---|
| Log of knowledge stock ($\beta_1$) | 0.607** | 0.613** | 0.713** | 0.472 |
| | (0.011) | (0.008) | (0.218) | (0.234) |
| Controls: | | | | |
| Weather | Yes | No | Yes | Yes |
| Trend | No | No | Linear | Hamilton filter |
| Degrees of freedom | 48 | 56 | 47 | 40 |
| Observations | 57 | 58 | 57 | 53 |

Notes: This is similar to Table S1 but the TFP and public R&D data are obtained from InSTePP. As a result, the number of observations is slightly smaller due to the shorter time series. These models are based on the best-fitting measure of the knowledge stock variable corresponding to research lags defined with gamma distribution parameters $\lambda = 0.4$ and $\delta = 0.95$ over 50 years. Standard errors are in parenthesis and are computed based on an overlapping block bootstrap to account to serial correlation (block size = 5 obs). ** $p < 0.01$, * $p < 0.05$. See SM for more details.

**Table S4.**

**Contribution of the knowledge stock and $CO_2$ concentrations to TFP growth**

| Dep. Variable: log($TFP$) | (1) | (2) | (3) | (4) | (5) | (6) |
|---|---|---|---|---|---|---|
| Log of knowledge stock | 0.503** | | 0.498** | 0.474** | | 0.748** |
| | (0.013) | | (0.038) | (0.154) | | (0.149) |
| $CO_2$ (June–August) | | 0.010** | 0.000 | | –0.001 | 0.001 |
| | | (0.001) | (0.001) | | (0.002) | (0.001) |
| Controls: | | | | | | |
|   Weather | Yes | Yes | Yes | Yes | Yes | Yes |
|   Trend | No | No | No | Hamilton filter | Hamilton filter | Hamilton filter |
| Degrees of freedom | 62 | 54 | 53 | 54 | 50 | 49 |
| Observations | 71 | 63 | 63 | 67 | 63 | 63 |

Notes: These models are based on the best-fitting measure of the knowledge stock variable corresponding to research lags defined with a gamma distribution with parameters $\lambda = 0.75$ and $\delta = 0.9$ over 50 years. Standard errors are in parenthesis and are computed based on an overlapping block bootstrap to account to serial correlation (block size = 5 obs). ** $p < 0.01$, * $p < 0.05$. See SM for more details.

**Table S5.**
**Projected impact of uniform changes in temperature on US agricultural TFP**

| Annual temperature change (Δ°C) | TFP Impact (%) | | |
|:---:|:---:|:---:|:---:|
| | $\mu$ | 2.5th | 97.5th |
| –3 | 3.8 | –4.3 | 11.1 |
| –2 | 3.3 | –2.1 | 8.2 |
| –1 | 2.0 | –0.6 | 4.6 |
| 0 | 0.0 | 0.0 | 0.0 |
| 1 | –2.8 | –5.4 | –0.2 |
| 2 | –6.5 | –11.7 | –1.2 |
| 3 | –10.8 | –19.0 | –2.9 |
| 4 | –15.8 | –27.1 | –4.8 |
| 5 | –21.3 | –35.4 | –7.5 |
| 6 | –27.2 | –43.7 | –10.1 |
| 7 | –33.2 | –51.8 | –12.6 |
| 8 | –39.2 | –59.7 | –16.0 |

Notes: $\mu$ refers to the mean estimate. The 2.5th and 97.5th columns correspond to percentile estimates obtained from bootstrapped estimates and represent a 95% confidence interval.

**Table S6.**

**Projected impact of uniform changes in precipitation on US agricultural TFP**

| Annual precipitation change (%) | TFP Impact (%) | | |
|---|---|---|---|
| | $\mu$ | 2.5th | 97.5th |
| –20 | –3.0 | –7.6 | 1.3 |
| –15 | –1.8 | –4.6 | 1.0 |
| –10 | –0.8 | –2.4 | 0.7 |
| –5 | –0.3 | –0.9 | 0.5 |
| 0 | 0.0 | 0.0 | 0.0 |
| +5 | –0.1 | –0.8 | 0.5 |
| +10 | –0.5 | –2.1 | 0.8 |
| +15 | –1.3 | –4.2 | 1.3 |
| +20 | –2.3 | –6.9 | 1.7 |

Notes: $\mu$ refers to the mean estimate. The 2.5th and 97.5th columns correspond to percentile estimates obtained from bootstrapped estimates and represent a 95% confidence interval.

**Table S7.**
**Projected impact of climate change on US agricultural TFP by 2025**

| GCM code | SSP1-2.6 | | | SSP2-4.5 | | | SSP3-7.0 | | | SSP5-8.5 | | |
|---|---|---|---|---|---|---|---|---|---|---|---|---|
| | $\mu$ | 2.5th | 97.5th | $\mu$ | 2.5th | 97.5th | $\mu$ | 2.5th | 97.5th | $\mu$ | 2.5th | 97.5th |
| ACCESS-CM2 | | | | -10.1 | -20.1 | -0.7 | -13.5 | -22.7 | -3.9 | | | |
| ACCESS-ESM1-5 | -2.0 | -7.9 | 5.2 | -8.9 | -16.1 | -1.6 | -7.0 | -13.2 | -1.5 | | | |
| BCC-CSM2-MR | -3.7 | -9.6 | 3.4 | -4.5 | -10.6 | 2.6 | -8.1 | -13.1 | -3.0 | -8.3 | -15.6 | -0.9 |
| CanESM5 | -8.7 | -16.3 | -1.3 | -3.7 | -9.8 | 3.1 | -3.2 | -10.8 | 5.9 | -10.1 | -18.5 | -0.8 |
| CMCC-CM2-SR5 | 1.7 | -5.4 | 10.8 | -2.7 | -7.6 | 2.0 | -1.3 | -7.1 | 6.7 | 0.8 | -5.3 | 8.9 |
| CMCC-ESM2 | -2.7 | -7.9 | 2.5 | -6.0 | -10.2 | -2.1 | -1.0 | -5.5 | 4.7 | -1.9 | -6.4 | 2.9 |
| EC-Earth3 | -6.2 | -12.5 | 1.0 | -3.1 | -7.9 | 2.2 | -7.1 | -12.9 | -1.2 | -2.3 | -7.7 | 4.7 |
| EC-Earth3-AerChem | | | | | | | -3.0 | -8.2 | 3.2 | | | |
| EC-Earth3-CC | | | | -5.0 | -11.4 | 2.3 | | | | -5.1 | -10.5 | 1.2 |
| EC-Earth3-Veg | -4.4 | -10.3 | 2.7 | -2.1 | -7.8 | 2.7 | -2.5 | -8.1 | 4.0 | -5.6 | -9.9 | -1.2 |
| EC-Earth3-Veg-LR | -4.2 | -8.1 | 0.1 | -4.6 | -8.1 | -1.3 | -12.8 | -20.6 | -3.2 | -3.0 | -5.9 | -0.1 |
| FGOALS-g3 | | | | -2.0 | -7.7 | 4.6 | -0.9 | -6.7 | 5.5 | -4.0 | -9.9 | 1.3 |
| GFDL-CM4 | | | | -9.1 | -16.3 | -2.2 | | | | -6.0 | -13.2 | 2.0 |
| GFDL-ESM4 | -11.0 | -17.5 | -5.0 | -5.9 | -10.9 | -1.7 | -7.4 | -12.3 | -1.8 | -3.9 | -7.9 | 0.4 |
| HadGEM3-GC31-LL | -10.4 | -17.7 | -3.5 | -10.5 | -18.1 | -3.0 | | | | -12.3 | -21.7 | -3.8 |
| INM-CM4-8 | -9.5 | -16.4 | -2.7 | -5.5 | -9.5 | -1.5 | -3.7 | -7.0 | -0.2 | -2.5 | -5.7 | 1.3 |
| INM-CM5-0 | -4.1 | -8.2 | -0.9 | -5.5 | -10.8 | -1.0 | -4.3 | -7.6 | -0.6 | -0.2 | -6.4 | 7.6 |
| MIROC-ES2L | -6.0 | -13.2 | 0.0 | -3.7 | -8.5 | 0.6 | -4.3 | -9.4 | 0.1 | -0.8 | -6.4 | 5.9 |
| MIROC6 | -3.7 | -7.9 | 0.0 | | | | -3.3 | -7.9 | 1.4 | -3.1 | -9.7 | 4.9 |
| MPI-ESM1-2-HR | -8.1 | -14.6 | -2.4 | | | | | | | -1.4 | -5.0 | 2.6 |
| MPI-ESM1-2-LR | 0.8 | -3.8 | 5.8 | -0.3 | -4.3 | 4.0 | -0.4 | -4.0 | 3.6 | -3.3 | -6.5 | -0.1 |
| NESM3 | -1.4 | -4.7 | 2.0 | -5.9 | -10.6 | 0.0 | | | | -5.9 | -12.0 | 0.1 |
| NorESM2-LM | -5.3 | -9.0 | -1.4 | -9.7 | -15.0 | -4.2 | -11.0 | -16.7 | -4.7 | -7.3 | -14.1 | -0.9 |
| NorESM2-MM | -4.2 | -8.7 | 0.1 | -3.4 | -7.6 | 0.5 | -2.2 | -9.2 | 6.5 | -7.2 | -12.5 | -2.0 |
| UKESM1-0-LL | -8.7 | -16.0 | -0.6 | -10.4 | -18.3 | -1.6 | -13.4 | -23.5 | -3.8 | -5.8 | -12.3 | 1.0 |
| Ensemble mean | -5.1 | -10.8 | 0.8 | -5.6 | -11.2 | 0.2 | -5.5 | -11.3 | 0.9 | -4.5 | -10.1 | 1.6 |

Notes: $\mu$ refers to the mean estimate. The 2.5th and 97.5th columns correspond to percentile estimates obtained from bootstrapped estimates and represent a 95% confidence interval.

**Table S8.**
**Projected impact of climate change on US agricultural TFP by 2050**

| GCM code | SSP1-2.6 | | | SSP2-4.5 | | | SSP3-7.0 | | | SSP5-8.5 | | |
|---|---|---|---|---|---|---|---|---|---|---|---|---|
| | $\mu$ | 2.5th | 97.5th | $\mu$ | 2.5th | 97.5th | $\mu$ | 2.5th | 97.5th | $\mu$ | 2.5th | 97.5th |
| ACCESS-CM2 | | | | -13.5 | -25.3 | -0.7 | -14.2 | -25.3 | -1.4 | | | |
| ACCESS-ESM1-5 | -6.1 | -14.8 | 2.8 | -14.4 | -25.2 | -3.4 | -11.7 | -20.7 | -2.5 | | | |
| BCC-CSM2-MR | -9.9 | -17.0 | -2.5 | -9.9 | -19.2 | 0.6 | -13.3 | -21.9 | -4.3 | -16.8 | -27.0 | -4.9 |
| CanESM5 | -19.2 | -31.0 | -7.8 | -17.1 | -29.4 | -6.0 | -15.8 | -26.6 | -3.2 | -18.3 | -31.0 | -6.0 |
| CMCC-CM2-SR5 | -3.5 | -10.9 | 5.8 | -7.9 | -17.1 | 0.7 | -10.3 | -20.7 | 1.0 | -11.2 | -21.1 | -1.3 |
| CMCC-ESM2 | -4.3 | -10.6 | 3.4 | -6.9 | -14.8 | 1.7 | -5.8 | -13.3 | 2.4 | -10.4 | -19.9 | 0.0 |
| EC-Earth3 | -3.9 | -12.5 | 4.7 | -11.8 | -21.5 | -2.2 | -13.2 | -22.8 | -4.6 | -14.4 | -24.8 | -2.6 |
| EC-Earth3-AerChem | | | | | | | -8.9 | -16.3 | -0.5 | | | |
| EC-Earth3-CC | | | | -4.3 | -11.3 | 3.7 | | | | -10.1 | -18.7 | -0.8 |
| EC-Earth3-Veg | -8.8 | -15.8 | -1.0 | -5.5 | -10.6 | 0.4 | -8.1 | -16.3 | 1.2 | -12.3 | -21.7 | -2.3 |
| EC-Earth3-Veg-LR | -7.3 | -12.4 | -1.5 | -8.4 | -19.9 | 2.5 | -9.3 | -15.5 | -2.2 | -14.4 | -24.1 | -5.0 |
| FGOALS-g3 | | | | -0.8 | -6.8 | 6.6 | -5.3 | -12.8 | 3.0 | -0.1 | -7.1 | 8.4 |
| GFDL-CM4 | | | | -7.4 | -15.5 | 1.3 | | | | -15.2 | -25.9 | -4.3 |
| GFDL-ESM4 | -7.5 | -13.9 | -1.9 | -7.9 | -13.9 | -1.9 | -6.5 | -12.4 | -0.9 | -18.0 | -28.5 | -8.0 |
| HadGEM3-GC31-LL | -17.6 | -28.2 | -7.3 | -10.1 | -20.0 | 0.4 | | | | -26.2 | -41.2 | -10.0 |
| INM-CM4-8 | -8.3 | -13.9 | -2.9 | -6.8 | -12.2 | -1.3 | -9.7 | -16.5 | -2.3 | -15.2 | -24.9 | -6.0 |
| INM-CM5-0 | -9.3 | -16.8 | -3.0 | -9.4 | -17.2 | -2.8 | -5.7 | -12.4 | 1.2 | -15.2 | -26.3 | -5.2 |
| MIROC-ES2L | -10.6 | -23.2 | 1.4 | -11.5 | -20.2 | -4.5 | -12.1 | -21.2 | -4.1 | -16.0 | -27.5 | -5.5 |
| MIROC6 | -3.7 | -11.3 | 5.1 | | | | -9.0 | -16.9 | -0.6 | -7.2 | -13.8 | -0.2 |
| MPI-ESM1-2-HR | -6.5 | -11.7 | -0.9 | | | | | | | -8.1 | -14.2 | -1.8 |
| MPI-ESM1-2-LR | -6.5 | -11 | -1.6 | -4.8 | -10.0 | 0.9 | -12.3 | -20.9 | -4.7 | -2.5 | -7.0 | 2.7 |
| NESM3 | 1.0 | -4.4 | 7.7 | -2.3 | -9.5 | 5.9 | | | | -11.0 | -19.8 | -1.6 |
| NorESM2-LM | -6.5 | -13.5 | 0.6 | -8.6 | -17.1 | 0.4 | -6.0 | -12.1 | 0.1 | -7.7 | -16.5 | 1.5 |
| NorESM2-MM | -7.4 | -12.7 | -1.9 | -9.3 | -17.8 | -1.4 | -6.3 | -13.4 | 1.3 | -14.4 | -23.5 | -5.6 |
| UKESM1-0-LL | -9.3 | -19.4 | 2.3 | -13 | -23.9 | -1.7 | -20.8 | -34.3 | -7.2 | -24.0 | -39.2 | -8.5 |
| Ensemble mean | -7.8 | -15.3 | 0.1 | -8.7 | -17.2 | 0.0 | -10.2 | -18.6 | -1.4 | -13.1 | -22.9 | -3.1 |

Notes: $\mu$ refers to the mean estimate. The 2.5th and 97.5th columns correspond to percentile estimates obtained from bootstrapped estimates and represent a 95% confidence interval.

**Table S9.**
**Projected impact of climate change on US agricultural TFP by 2075**

| GCM code | SSP1-2.6 | | | SSP2-4.5 | | | SSP3-7.0 | | | SSP5-8.5 | | |
|---|---|---|---|---|---|---|---|---|---|---|---|---|
| | $\mu$ | 2.5th | 97.5th | $\mu$ | 2.5th | 97.5th | $\mu$ | 2.5th | 97.5th | $\mu$ | 2.5th | 97.5th |
| ACCESS-CM2 | | | | -20.5 | -33.4 | -7.1 | -25.3 | -40.9 | -8.2 | | | |
| ACCESS-ESM1-5 | -11.1 | -21.1 | -0.5 | -20.2 | -31.6 | -8.0 | -14.5 | -26.8 | -1.4 | | | |
| BCC-CSM2-MR | -8.0 | -14.8 | -1.3 | -11.8 | -20.9 | -2.4 | -21.4 | -35.2 | -7.3 | -24.7 | -39.6 | -9.4 |
| CanESM5 | -11.7 | -21.8 | -1.3 | -18.2 | -31.7 | -4.6 | -24.0 | -39.4 | -7.8 | -33.1 | -51.6 | -12.0 |
| CMCC-CM2-SR5 | -7.6 | -17.0 | 1.3 | -9.9 | -19.9 | 1.1 | -11.0 | -21.9 | 0.9 | -20.6 | -36.0 | -5.4 |
| CMCC-ESM2 | -9.8 | -20.8 | 3.1 | -10.0 | -21.7 | 2.9 | -17.7 | -31.2 | -4.1 | -18.9 | -33.0 | -2.5 |
| EC-Earth3 | -7.0 | -13.4 | 0.0 | -14.4 | -25.6 | -3.7 | -14.4 | -24.9 | -2.3 | -22.4 | -36.1 | -6.1 |
| EC-Earth3-AerChem | | | | | | | -26.5 | -42.4 | -10.5 | | | |
| EC-Earth3-CC | | | | -11.5 | -21.1 | -3.1 | | | | -25.2 | -41.2 | -10.1 |
| EC-Earth3-Veg | -5.2 | -12.3 | 3.1 | -12.9 | -22.6 | -2.7 | -13.0 | -23.7 | -1.4 | -17.0 | -29.4 | -3.1 |
| EC-Earth3-Veg-LR | -9.5 | -18.0 | -1.0 | -15.1 | -24.9 | -5.3 | -16.0 | -28.1 | -3.6 | -22.3 | -36.1 | -8.6 |
| FGOALS-g3 | | | | -8.3 | -16.7 | 0.1 | -8.8 | -17.6 | 0.4 | -11.0 | -20.1 | -1.3 |
| GFDL-CM4 | | | | -14.3 | -24.0 | -4.6 | | | | -30.7 | -46.3 | -13.6 |
| GFDL-ESM4 | -5.0 | -9.5 | -0.4 | -8.2 | -16.2 | 1.0 | -11.9 | -22.0 | -1.2 | -13.1 | -23.0 | -3.1 |
| HadGEM3-GC31-LL | -17.3 | -28.8 | -5.4 | -23.1 | -37.9 | -8.1 | | | | -34.5 | -52.9 | -14.6 |
| INM-CM4-8 | -8.1 | -14.7 | -1.3 | -8.4 | -16.3 | 0.8 | -9.7 | -17.6 | -0.9 | -17.9 | -29.2 | -6.2 |
| INM-CM5-0 | -5.4 | -11.7 | 0.0 | -6.7 | -12.0 | -0.6 | -11.1 | -19.3 | -3.3 | -11.8 | -21.0 | -2.6 |
| MIROC-ES2L | -6.8 | -13.7 | 0.0 | -19.9 | -33.2 | -7.9 | -15.9 | -29.1 | -1.5 | -25.3 | -42.1 | -8.1 |
| MIROC6 | -5.9 | -11.9 | 0.1 | | | | -16.2 | -28.0 | -4.6 | -19.2 | -33.0 | -5.1 |
| MPI-ESM1-2-HR | -3.3 | -6.8 | 0.4 | | | | | | | -18.6 | -30.8 | -6.1 |
| MPI-ESM1-2-LR | 0.0 | -3.7 | 4.1 | -3.5 | -8.1 | 2.0 | -19.2 | -31.0 | -7.4 | -15.1 | -25.7 | -3.6 |
| NESM3 | -7.6 | -17.7 | 0.6 | -8.8 | -18.1 | 0.5 | | | | -22.3 | -35.3 | -7.1 |
| NorESM2-LM | -9.6 | -17.0 | -2.1 | -5.4 | -12.3 | 2.0 | -16.7 | -28.2 | -5.3 | -16.0 | -28.3 | -3.2 |
| NorESM2-MM | -5.5 | -12.6 | 2.2 | -9.5 | -18.4 | -0.6 | -16.9 | -29.0 | -2.7 | -13.3 | -24.9 | -0.5 |
| UKESM1-0-LL | -10.3 | -20.2 | 0.8 | -18.6 | -31.7 | -4.4 | -35.2 | -54.1 | -14.9 | -41.3 | -61.4 | -17.9 |
| Ensemble mean | -7.7 | -15.4 | 0.1 | -12.7 | -22.7 | -2.4 | -17.3 | -29.5 | -4.4 | -21.6 | -35.3 | -6.8 |

Notes: $\mu$ refers to the mean estimate. The 2.5th and 97.5th columns correspond to percentile estimates obtained from bootstrapped estimates and represent a 95% confidence interval.

**Table S10.**
**Projected impact of climate change on US agricultural TFP by 2100**

| GCM code | SSP1-2.6 | | | SSP2-4.5 | | | SSP3-7.0 | | | SSP5-8.5 | | |
|---|---|---|---|---|---|---|---|---|---|---|---|---|
| | $\mu$ | 2.5th | 97.5th | $\mu$ | 2.5th | 97.5th | $\mu$ | 2.5th | 97.5th | $\mu$ | 2.5th | 97.5th |
| ACCESS-CM2 | | | | -25.8 | -41.2 | -9.5 | -34.1 | -51.5 | -12.4 | | | |
| ACCESS-ESM1-5 | -3.0 | -12.3 | 9.1 | -12.6 | -24.3 | 0.4 | -23.4 | -38.7 | -7.2 | | | |
| BCC-CSM2-MR | -4.9 | -10.3 | 1.1 | -14.7 | -24.6 | -5.0 | -34.1 | -50.6 | -14.3 | -37.3 | -56.4 | -16.9 |
| CanESM5 | -6.7 | -15.3 | 2.6 | -14.0 | -26.1 | -0.9 | -36.0 | -54.5 | -14.4 | -44.6 | -65.4 | -19.4 |
| CMCC-CM2-SR5 | -19.4 | -32.4 | -6.3 | -9.8 | -21.8 | 4.4 | -25.2 | -40.1 | -6.5 | -36.0 | -55.7 | -13.6 |
| CMCC-ESM2 | -10.9 | -22.1 | 0.5 | -19.8 | -34.1 | -5.3 | -18.8 | -33.7 | -2.6 | -34.4 | -53.0 | -11.6 |
| EC-Earth3 | -6.0 | -12.4 | 1.8 | -19.6 | -32.8 | -7.1 | -38.2 | -57.1 | -17.5 | -40.0 | -59.4 | -17.1 |
| EC-Earth3-AerChem | | | | | | | -35.6 | -53.4 | -13.1 | | | |
| EC-Earth3-CC | | | | -15.0 | -26.3 | -3.2 | | | | -42.5 | -63.1 | -18.6 |
| EC-Earth3-Veg | -7.3 | -14.2 | -0.2 | -14.6 | -24.9 | -2.6 | -24.0 | -39.4 | -6.9 | -34.5 | -52.9 | -14.0 |
| EC-Earth3-Veg-LR | -7.9 | -14.1 | -2.2 | -17.6 | -35.7 | 0.3 | -28.1 | -43.9 | -11.2 | -41.4 | -62.5 | -18.1 |
| FGOALS-g3 | | | | -13.7 | -23.7 | -5.4 | -12.4 | -23.4 | -0.5 | -16.2 | -29.0 | -1.5 |
| GFDL-CM4 | | | | -21.2 | -34.1 | -7.5 | | | | -39.8 | -60.0 | -16.8 |
| GFDL-ESM4 | -7.2 | -15.4 | 1.0 | -12.2 | -21.2 | -3.5 | -23.2 | -37.7 | -8.4 | -25.5 | -41.4 | -9.5 |
| HadGEM3-GC31-LL | -21.4 | -34.5 | -8.2 | -28.8 | -44.9 | -11.3 | | | | -55.9 | -78.3 | -27.0 |
| INM-CM4-8 | -8.5 | -14.5 | -2.6 | -12.9 | -22.0 | -3.8 | -14.7 | -26.9 | -1.0 | -31.4 | -48.7 | -12.5 |
| INM-CM5-0 | -1.4 | -5.2 | 2.2 | -10.7 | -18.8 | -2.9 | -26.7 | -43.0 | -11.2 | -23.9 | -38.6 | -8.4 |
| MIROC-ES2L | -7.5 | -15.9 | 1.0 | -18.2 | -31.7 | -4.2 | -28.7 | -44.1 | -10.0 | -31.9 | -48.6 | -10.3 |
| MIROC6 | -11.8 | -19.8 | -3.7 | | | | -23.5 | -38.9 | -7.5 | -31.2 | -47.5 | -10.3 |
| MPI-ESM1-2-HR | -9.9 | -16.3 | -3.6 | | | | | | | -28.4 | -44.6 | -11.0 |
| MPI-ESM1-2-LR | -1.6 | -7.6 | 5.4 | -12.0 | -20.6 | -3.4 | -34.8 | -53.4 | -16.1 | -30.5 | -47.3 | -12.2 |
| NESM3 | -7.4 | -18.9 | 3.4 | -6.3 | -13.4 | 1.8 | | | | -30.6 | -47.1 | -11.3 |
| NorESM2-LM | -5.7 | -11.8 | 0.7 | -9.6 | -18.3 | 0.3 | -24.1 | -39.0 | -8.5 | -35.5 | -53.8 | -16.0 |
| NorESM2-MM | -7.1 | -14.3 | 0.6 | -13.0 | -24.0 | -1.8 | -23.7 | -38.2 | -8.8 | -31.0 | -48.4 | -12.1 |
| UKESM1-0-LL | -12.4 | -22.4 | -2.4 | -26.2 | -42.1 | -9.5 | -43.7 | -63.7 | -20.2 | -51.7 | -73.1 | -23.1 |
| Ensemble mean | -8.4 | -16.5 | 0.0 | -15.8 | -27.6 | -3.6 | -27.6 | -43.6 | -9.9 | -35.2 | -53.4 | -14.2 |

Notes: $\mu$ refers to the mean estimate. The 2.5th and 97.5th columns correspond to percentile estimates obtained from bootstrapped estimates and represent a 95% confidence interval.

**Table S11.**
**Change in the knowledge stock needed to offset climate change impacts on US agricultural TFP over time.**

|  |  | 2025 | 2050 | 2075 | 2100 |
|---|---|---|---|---|---|
|  |  | Relative change in the R&D stock (%) | | | |
| SSP1-2.6 | $\mu$ | 9.5 | 16.7 | 18.3 | 16.0 |
|  | 2.5th | -0.6 | -0.1 | 0.1 | -0.9 |
|  | 97.5th | 23.9 | 43.9 | 49.1 | 48.5 |
| SSP2-4.5 | $\mu$ | 10.6 | 22.3 | 33.2 | 43.2 |
|  | 2.5th | -0.3 | 2.7 | 5.7 | 6.7 |
|  | 97.5th | 26.7 | 59.2 | 94.7 | 128.0 |
| SSP3-7.0 | $\mu$ | 9.0 | 23.5 | 51.3 | 93.9 |
|  | 2.5th | -1.4 | 2.5 | 9.8 | 17.8 |
|  | 97.5th | 24.4 | 64.3 | 152.5 | 336.8 |
| SSP5-8.5 | $\mu$ | 9.3 | 28.2 | 72.3 | 151.0 |
|  | 2.5th | -1.5 | 4.2 | 14.4 | 28.8 |
|  | 97.5th | 25.8 | 79.1 | 239.3 | 671.3 |
|  |  | Absolute change in the R&D stock ($B) | | | |
| SSP1-2.6 | $\mu$ | 0.589 | 0.912 | 0.938 | 0.823 |
|  | 2.5th | -0.037 | -0.003 | 0.003 | -0.042 |
|  | 97.5th | 1.450 | 2.322 | 2.435 | 2.408 |
| SSP2-4.5 | $\mu$ | 0.657 | 1.223 | 1.730 | 2.292 |
|  | 2.5th | -0.017 | 0.144 | 0.283 | 0.335 |
|  | 97.5th | 1.620 | 3.132 | 4.697 | 6.351 |
| SSP3-7.0 | $\mu$ | 0.558 | 1.295 | 2.740 | 5.321 |
|  | 2.5th | -0.083 | 0.131 | 0.488 | 0.881 |
|  | 97.5th | 1.479 | 3.401 | 7.561 | 16.705 |
| SSP5-8.5 | $\mu$ | 0.575 | 1.566 | 3.995 | 9.417 |
|  | 2.5th | -0.089 | 0.221 | 0.716 | 1.426 |
|  | 97.5th | 1.562 | 4.186 | 11.867 | 33.293 |

Notes: $\mu$ refers to the mean estimate. The 2.5th and 97.5th rows correspond to percentile estimates obtained from bootstrapped estimates and represent a 95% confidence interval. The top and bottom portions of the table correspond to the first and second row of panels in Fig. S16, respectively.

**Table S12.**
**Annual R&D spending growth (%) needed starting in 2021 to offset climate change impacts on US agricultural TFP by certain target years.**

|  | 2050 | 2060 | 2070 | 2080 | 2090 | 2100 |
|---|---|---|---|---|---|---|
|  | | | SSP1-2.6 | | | |
| $\mu$ | 5.163 | 1.788 | 0.938 | 0.588 | 0.413 | 0.313 |
| 2.5th | -0.012 | 0.012 | 0.012 | -0.012 | -0.012 | -0.012 |
| 97.5th | 10.163 | 3.888 | 2.113 | 1.388 | 1.038 | 0.813 |
|  | | | SSP2-4.5 | | | |
| $\mu$ | 6.487 | 2.513 | 1.513 | 1.112 | 0.913 | 0.788 |
| 2.5th | 0.988 | 0.413 | 0.262 | 0.213 | 0.163 | 0.137 |
| 97.5th | 12.238 | 5.288 | 3.263 | 2.413 | 1.938 | 1.688 |
|  | | | SSP3-7.0 | | | |
| $\mu$ | 6.763 | 3.013 | 2.062 | 1.713 | 1.563 | 1.513 |
| 2.5th | 0.913 | 0.588 | 0.438 | 0.388 | 0.338 | 0.338 |
| 97.5th | 12.838 | 6.112 | 4.288 | 3.513 | 3.238 | 3.013 |
|  | | | SSP5-8.5 | | | |
| $\mu$ | 7.788 | 3.688 | 2.688 | 2.338 | 2.213 | 2.188 |
| 2.5th | 1.488 | 0.838 | 0.613 | 0.538 | 0.513 | 0.513 |
| 97.5th | 14.413 | 7.413 | 5.438 | 4.688 | 4.363 | 4.188 |

Notes: $\mu$ refers to the mean estimate. The 2.5th and 97.5th rows correspond to percentile estimates obtained from bootstrapped estimates and represent a 95% confidence interval.

**Table S13.**
**Cumulative R&D spending (2020 $B) needed starting in 2021 to offset climate change impacts on US agricultural TFP by certain target years.**

|  | 2050 | 2060 | 2070 | 2080 | 2090 | 2100 |
|---|---|---|---|---|---|---|
| | | | SSP1-2.6 | | | |
| $\mu$ | 207.6 | 92.9 | 69.5 | 60.1 | 56.0 | 54.6 |
| 2.5th | -0.3 | 0.5 | 0.8 | -1.1 | -1.5 | -2.0 |
| 97.5th | 778.1 | 278.4 | 194.1 | 168.5 | 164.6 | 163.6 |
| | | | SSP2-4.5 | | | |
| $\mu$ | 306.4 | 145.3 | 124.3 | 127.1 | 140.2 | 157.4 |
| 2.5th | 25.1 | 17.7 | 17.3 | 20.1 | 20.8 | 22.9 |
| 97.5th | 1258.0 | 478.5 | 376.6 | 372.0 | 391.7 | 444.3 |
| | | | SSP3-7.0 | | | |
| $\mu$ | 330.5 | 187.9 | 187.7 | 223.8 | 284.8 | 376.6 |
| 2.5th | 23.0 | 25.8 | 29.8 | 38.0 | 45.0 | 59.4 |
| 97.5th | 1440.8 | 639.5 | 615.6 | 716.1 | 966.0 | 1255.9 |
| | | | SSP5-8.5 | | | |
| $\mu$ | 433.7 | 255.8 | 276.2 | 353.9 | 483.8 | 679.9 |
| 2.5th | 39.8 | 38.1 | 42.9 | 54.4 | 71.3 | 94.7 |
| 97.5th | 2047.5 | 985.0 | 1013.6 | 1321.9 | 1891.3 | 2765.2 |

Notes: $\mu$ refers to the mean estimate. The 2.5th and 97.5th rows correspond to percentile estimates obtained from bootstrapped estimates and represent a 95% confidence interval.